\documentclass[letterpaper,journal]{IEEEtran}

\usepackage{amsmath,amsfonts,amssymb}
\usepackage{algorithm}
\usepackage{algpseudocode}
\usepackage{array}
\usepackage{booktabs}
\usepackage{makecell}
\usepackage{multirow}
\usepackage[caption=false,font=normalsize,labelfont=sf,textfont=sf]{subfig}
\usepackage{textcomp}
\usepackage{stfloats}
\usepackage{url}
\usepackage{verbatim}
\usepackage{graphicx}
\usepackage{cite}
\usepackage[table]{xcolor}
\usepackage{pifont}
\usepackage{upgreek}
\usepackage[hidelinks,hypertexnames=false]{hyperref}

\newcommand{\cmark}{\textcolor{green!50!black}{\ding{51}}}
\newcommand{\xmark}{\textcolor{red!70!black}{\ding{55}}}

\newcommand{\tablehead}[1]{\makecell[c]{#1}}
\definecolor{TableMetric}{gray}{0.95}
\definecolor{TableOurs}{gray}{0.92}

\newlength{\gateDiagHeight}

\begin{document}

\title{Heterogeneous 2D/1D Signal Representation Fusion for Underwater Acoustic Modulation Recognition Under Distribution Shift}

\author{Ronglai Qian, Liang An, Xiaoyan Wang, Qing Fan, Ziwei Huang, and Yang Ye%
\thanks{This work was supported by the Fundamental Research Funds for the Central Universities (No.~2242025F20003). (\textit{Corresponding author: Xiaoyan Wang.})}
\thanks{Ronglai Qian, Liang An, Qing Fan, Ziwei Huang, Yang Ye, and Xiaoyan Wang are with the Key Laboratory of Underwater Acoustic Signal Processing of the Ministry of Education, Southeast University, Nanjing 210096, China.(e-mail:\{
qianronglai, anliang, xyan, fanqing, huangziwei, yeyang1120\}@seu.edu.cn)}
}

\markboth{IEEE Transactions on Image Processing}%
{Qian \MakeLowercase{\textit{et al.}}: Underwater Acoustic Modulation Recognition Under Distribution Shift}

\maketitle

\begin{abstract}
Modulation recognition systems rely on heterogeneous signal representations. 2D signal-image modalities such as time-frequency and cyclostationary maps capture structural patterns, while 1D statistical descriptors such as higher-order power spectra encode complementary cues. Under distribution shift, these modalities degrade unevenly, making robust fusion a central challenge for practical deployment. Progress is further limited by the lack of a unified evaluation protocol that systematically separates different shift types. This paper addresses both challenges through a joint benchmark-and-model study in underwater acoustic modulation recognition. UAMR-ShiftBench is the first benchmark to jointly cover in-distribution, low-SNR, unseen-environment, unseen-communication-parameter, and measured sea-trial evaluation under a single matched protocol, with two independent real-world subsets collected during two sea-trial campaigns conducted in March and November in the South China Sea. SCP-TriCA fuses STFT, cyclostationary, and P2/P4 (second- and fourth-order power spectra)  modalities hierarchically: the two 2D modalities are first aligned through bidirectional cross-attention, and the 1D statistical modality is then incorporated through a sample-adaptive selective gate. On UAMR-ShiftBench, SCP-TriCA achieves 95.33\% in-distribution accuracy and 74.59\% simulated OOD average, outperforming the strongest baseline by 5.12 percentage points, and reaches 91.14\% and 94.86\% on the two sea-trial subsets, exceeding the best baseline by 15.71 and 23.00 percentage points respectively. Ablation results confirm that the gains stem from modality complementarity and the hierarchical fusion design.  Code and models are available at \url{https://github.com/ronglaiqian/UAMR-ShiftBench}.
\end{abstract}


\section{Introduction}\label{sec:introduction}

\IEEEPARstart{H}{eterogeneous} representations derived from raw waveforms are widely used to bridge signal analysis and modern recognition models. A received waveform can be described by 2D signal-image modalities, such as time-frequency or cyclostationary maps, together with 1D statistical descriptors, such as higher-order power spectra. These representations capture complementary aspects of the underlying signal, but their discriminative reliability does not degrade uniformly under distribution shift: a modality that is highly informative under matched conditions may become unreliable under noise corruption or domain gap, while another retains its structure. Designing fusion architectures that remain robust under such uneven reliability change is therefore a fundamental challenge for practical signal recognition systems \cite{HuynhThe2021,Antoni2007CyclicSpectralPractice,LiWu2019DenseFuse,LiuLiuJiangFanLuo2021BilevelFusion,ZhangChengTian2020MultiviewVSVC,YehHuangWang2014HeterogeneousDA}.

Non-cooperative underwater acoustic modulation recognition is a particularly demanding testbed for this problem. The task infers waveform classes from passive observations under realistic deployment conditions, where test signals routinely differ from training data due to reduced SNR, unseen propagation environments, and variation in communication parameters. More critically, measured sea-trial data are expensive to collect and typically unavailable during training, making the simulation-to-real gap the hardest and most practically consequential shift to overcome. This gap is rarely evaluated directly: most existing methods are assessed either under controlled simulated conditions or with limited transfer to a single sea trial, making it difficult to determine whether strong simulation performance generalizes to independent real-world recordings. Closing this gap requires not only a robust fusion architecture, but also an evaluation protocol that treats simulation-to-real transfer as a first-class test condition rather than an afterthought.

Two gaps limit progress toward robust heterogeneous representation learning in this setting. The first is an evaluation gap. Existing methods are typically assessed on datasets that differ in class taxonomy, channel settings, and preprocessing choices, making cross-paper comparisons unreliable and robustness claims difficult to verify. More importantly, different shift types, including low SNR, unseen propagation environments, unseen communication parameters, and simulation-to-real transfer, are rarely separated as distinct test conditions under a unified protocol~\cite{MorenoTorres2012DatasetShift,Zhang2025ReliableDG}. Without this separation, it is impossible to determine whether a reported gain reflects genuine robustness or adaptation to one particular test configuration. The second is a fusion gap. Although multimodal fusion has emerged as a promising direction ~\cite{ZhangFengXuLuZhou2018MultiReceiver,Li2024TSTR,Wang2024FeatureFusion,WangHuangShiMao2024}, existing strategies such as early concatenation or simultaneous feature mixing treat all modalities symmetrically and provide no mechanism for controlling how a degraded modality influences the fused representation. Under distribution shift, where individual modalities degrade unevenly, this limitation directly undermines robustness.

To address both gaps, this paper presents UAMR-ShiftBench and SCP-TriCA. To the best of our knowledge, UAMR-ShiftBench is the first unified benchmark for underwater acoustic modulation recognition that jointly covers in-distribution, low-SNR, unseen-environment, unseen-communication-parameter, and measured sea-trial evaluation under a single matched protocol for seven waveform classes. The key design principle is shift disentanglement: rather than merging different shift types into a single aggregate test set, UAMR-ShiftBench organizes them as separate evaluation conditions, so that performance degradation can be attributed to specific mismatch factors and results remain comparable across models. The benchmark further includes two independent measured sea-trial subsets collected from real shallow-water environments, providing a direct zero-shot simulation-to-real assessment that is absent from existing underwater acoustic evaluation settings.

\begin{figure*}[!t]
    \centering
    \subfloat[]{%
    \includegraphics[width=0.52\textwidth]{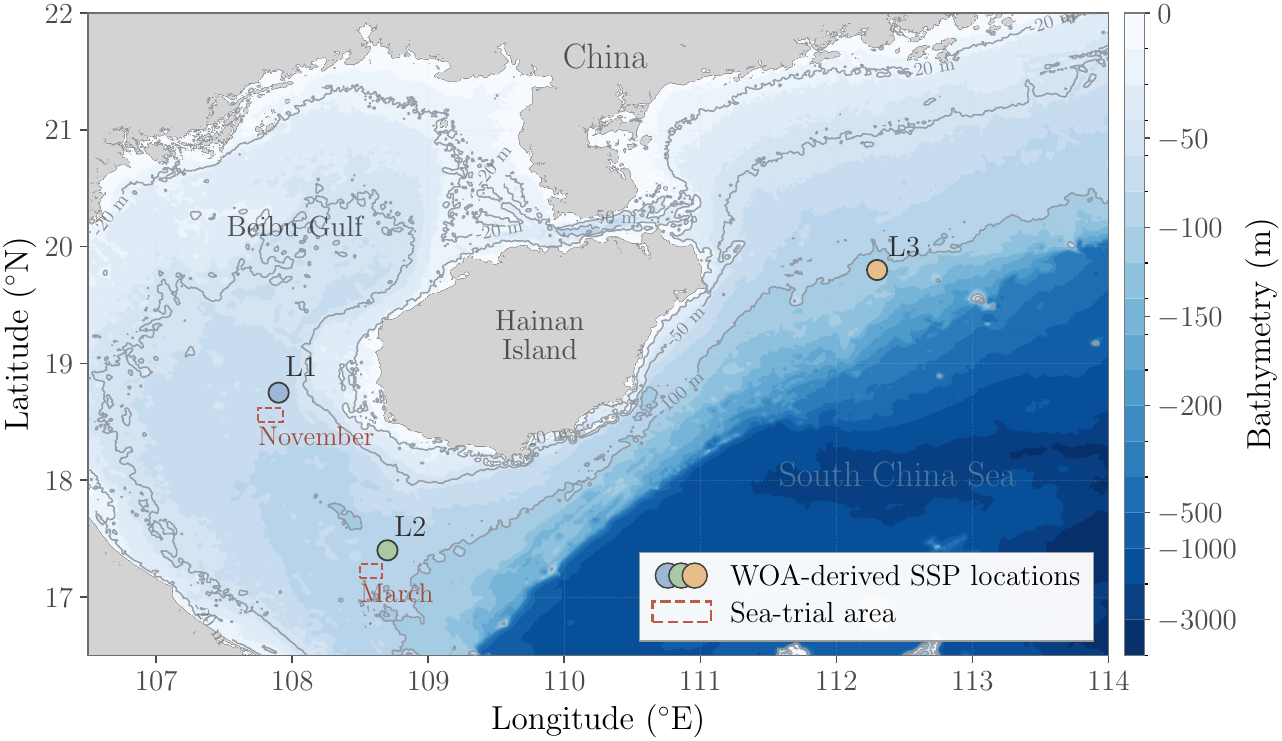}%
    \label{fig:benchmark_environment_a}}
    \hspace{0.03\textwidth}
    \subfloat[]{%
    \includegraphics[width=0.30\textwidth]{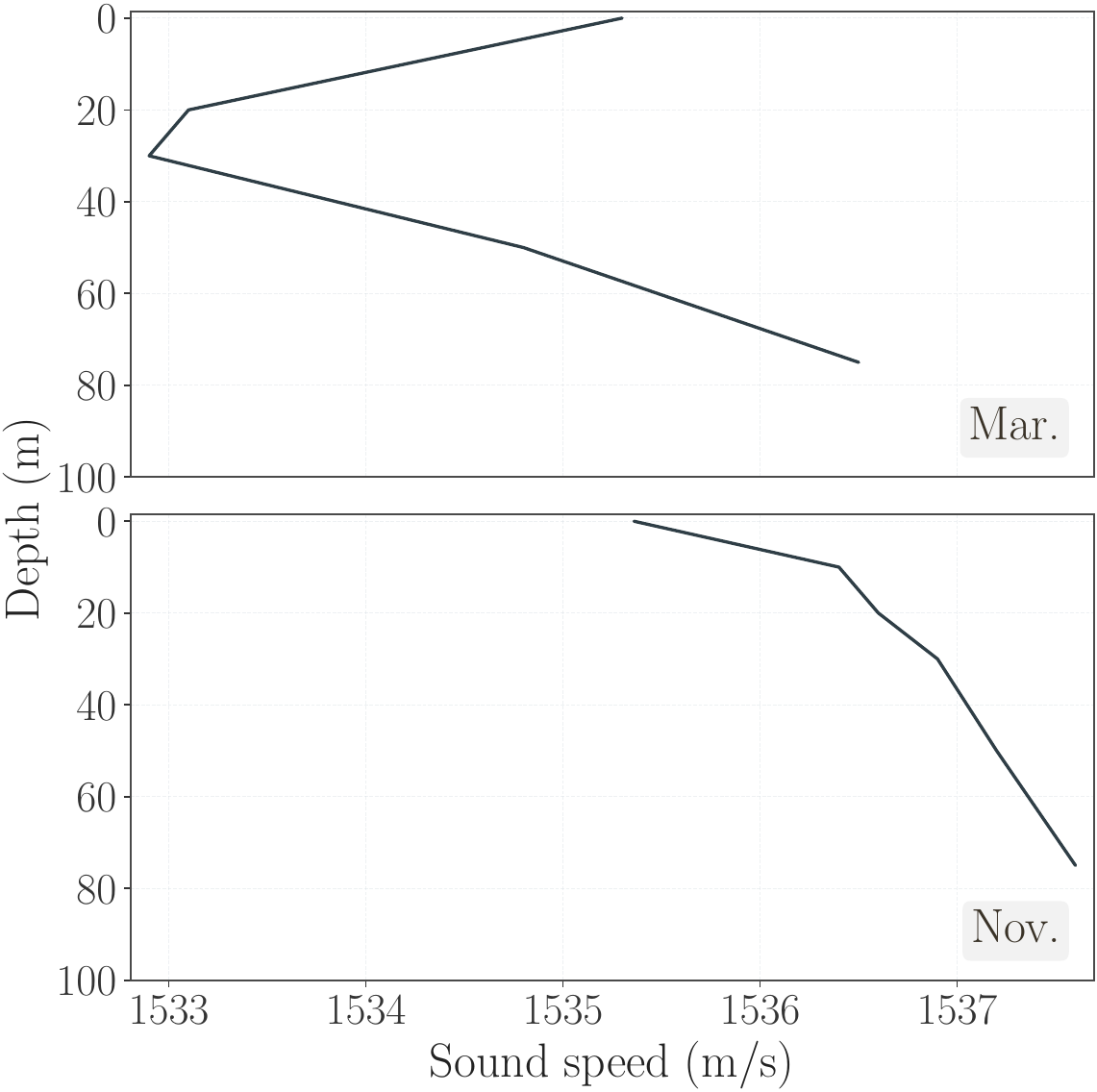}%
    \label{fig:benchmark_environment_b}}
    \par\vspace{-1em}
    \subfloat[]{%
    \includegraphics[width=0.86\textwidth]{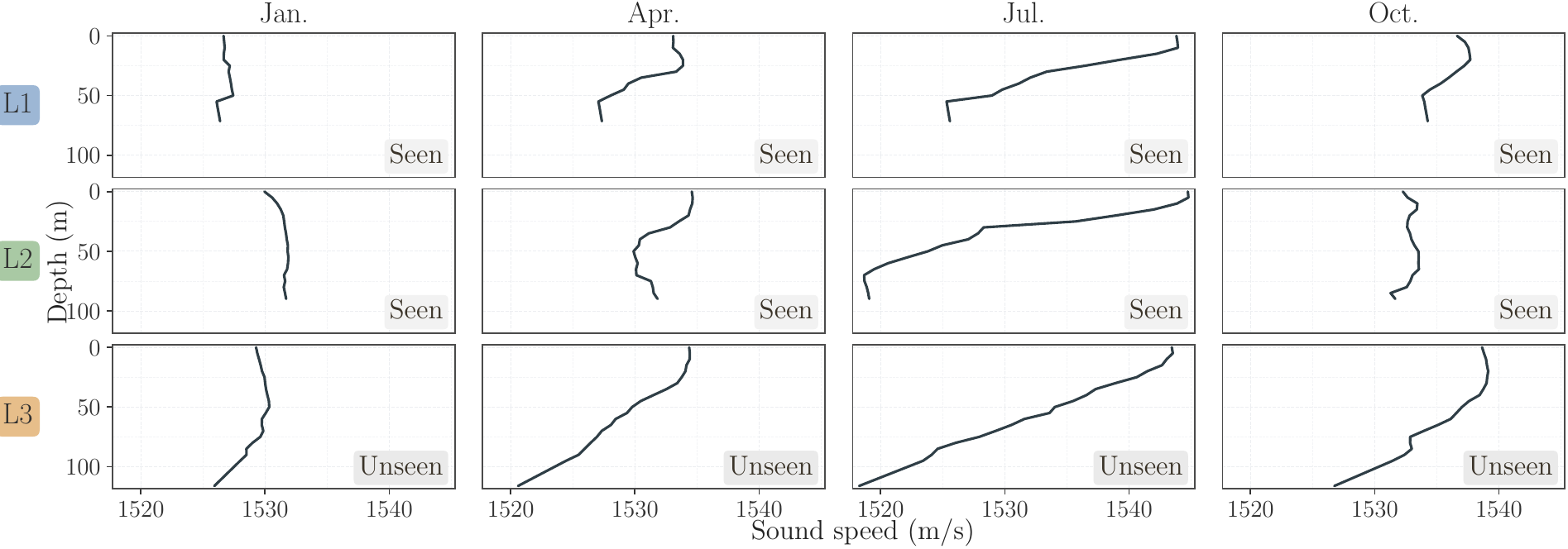}%
    \label{fig:benchmark_environment_c}}
    \caption{Environment configuration of UAMR-ShiftBench. (a) Study area and simulation locations, with bathymetric/topographic data from the GEBCO\_2025 Grid \cite{GEBCO2025}. (b) Sea-trial sound-speed profiles. (c) Monthly sound-speed profiles adopted for channel simulation.}
    \label{fig:benchmark_environment}
\end{figure*}

SCP-TriCA is a tri-modal cross-attention framework for robust heterogeneous 2D/1D signal representation fusion. The design is motivated by a simple observation: under distribution shift, the two 2D signal-image modalities and the 1D statistical modality do not degrade in the same way or at the same rate, and a fusion strategy that treats them symmetrically will be pulled off course by whichever modality happens to be most corrupted. SCP-TriCA addresses this by organizing fusion hierarchically. In the first stage, the STFT and cyclostationary modalities, which share the same 2D token geometry and jointly encode the dominant discriminative cues, are aligned through bidirectional cross-attention before being compressed into a unified 2D representation. In the second stage, this fused 2D representation queries the P2/P4 statistical token sequence through a second cross-attention block, and a sample-adaptive selective gate controls how much statistical evidence is incorporated. This gated injection allows the model to draw on higher-order statistical cues when they are informative, while protecting the fused 2D representation from corruption when the statistical branch is unreliable. Together, UAMR-ShiftBench and SCP-TriCA form a self-consistent study: the benchmark exposes the robustness problem in a controlled and reproducible way, and the model demonstrates that explicitly accounting for modality reliability under shift leads to measurable and consistent gains.

The main contributions of this paper are as follows:
\begin{itemize}
    \item We establish UAMR-ShiftBench, the first unified benchmark for underwater acoustic modulation recognition that jointly covers low-SNR, unseen-environment, unseen-communication-parameter, and simulation-to-real evaluation under a unified protocol, enabling systematic and fair comparison across practically relevant distribution shifts.
    \item We propose SCP-TriCA, a hierarchical tri-modal cross-attention framework that explicitly organizes heterogeneous 2D signal-image and 1D statistical evidence according to their changing reliability under distribution shift. On UAMR-ShiftBench, SCP-TriCA achieves the best simulated OOD average and the strongest sea-trial results among all evaluated methods, outperforming the strongest baseline by 5.12 percentage points in OOD average and by 15.71 and 23.00 percentage points on two independent March and November sea-trial subsets.
    \item We provide systematic ablation results demonstrating that the observed gains stem from modality complementarity and the proposed hierarchical fusion design, offering actionable insights into how heterogeneous signal representations should be organized for robust underwater acoustic recognition.
\end{itemize}

\section{Related Work}\label{sec:related_work}
\subsection{Signal Representations and Fusion for Modulation Recognition}\label{subsec:related_uamr}
Deep learning has moved modulation recognition from likelihood-based and feature-based pipelines toward models that learn discriminative representations from waveform sequences, time-frequency maps, higher-order statistics, or complex-valued signal forms. In underwater acoustic modulation recognition, representative studies have improved temporal modeling, local feature extraction, lightweight inference, high-order statistical characterization, and complex-valued representation learning through recurrent-convolutional networks, CNN variants, transformer-style backbones, and domain-specific feature encoders \cite{OSheaCorganClancy2016,HuynhThe2021,Zhang2022RCNN,Shao2025IQFormer,WangJinZhang2021,WangZhangXuCaoGulliver2020,HuangLiYangWang2022,ZhangHeJing2023,CuiZhengLiLiJiang2026WCCNet}. These single-view or dominant-view models can be effective under matched or moderately controlled conditions, but their performance may degrade when the relied-upon view is distorted by noise, multipath, environmental mismatch, or signal-parameter variation.

This limitation has motivated multi-view and multimodal recognition. Existing underwater studies exploit complementary information from receiver-side spatial diversity, sequence and time-frequency streams, multiple descriptive attributes, or richer multimodal transformations \cite{ZhangFengXuLuZhou2018MultiReceiver,Li2024TSTR,Wang2024FeatureFusion,WangHuangShiMao2024}. These efforts show that combining heterogeneous representations is a promising direction, especially when no single view captures all class-discriminative cues under practical channels. In these models, fusion is commonly implemented by concatenating features, mixing modality streams with shared blocks, or aggregating multiple representations in a single interaction stage. Such designs exploit complementarity, but they provide limited mechanisms for controlling how an unreliable view influences the final representation under distribution shift.

This fusion issue is also consistent with broader observations in image fusion and multi-view representation learning. Studies on infrared-visible fusion and multi-view image classification have shown that heterogeneous information can be integrated through learned fusion mappings, model-guided constraints, transformer-based cross-modal relation modeling, and view-consistency regularization \cite{LiWu2019DenseFuse,LiuLiuJiangFanLuo2021BilevelFusion,RaoXuWu2023TGFuse,ZhangChengTian2020MultiviewVSVC}. Although these studies are not designed for underwater acoustic modulation recognition, they provide useful methodological support for structured interaction among heterogeneous representations rather than naive feature concatenation. In this work, we instantiate this principle for signal-derived 2D/1D modalities, whose physical meanings and reliability profiles differ under underwater noise, propagation mismatch, and signal-parameter variation.

\subsection{Robust Recognition under Distribution Shift}\label{subsec:related_shift}
Robust recognition under distribution shift has received growing attention in wireless AMC, where channel mismatch, hardware impairment, sample scarcity, and deployment drift can seriously degrade performance. Representative efforts include multipath-aware evaluation, adversarial transfer learning, meta-learning, few-shot recognition, consistency regularization, and channel- or hardware-aware data augmentation \cite{TekbiyikAydinEkti2019,BuHeJingHan2020,HaoFengYang2023,Su2025FewShot,Yang2026Consistency,PerendaBovetZhelevaPollin2024}. Collectively, these studies move the goal of AMC beyond fitting a single benchmark toward learning representations that remain discriminative under changing operating conditions.

Distribution shift is particularly challenging in underwater acoustic modulation recognition because underwater propagation and ambient noise can vary strongly across environments, while non-cooperative transmitters may also use signal parameters that were not observed during training. These changes can alter waveform morphology, time-frequency structure, cyclostationary patterns, and higher-order statistical cues at the same time \cite{StojanovicPreisig2009,Antoni2007CyclicSpectralPractice}. Existing underwater studies have begun to address robustness through channel-estimation-assisted correction, multi-receiver fusion, complex-valued processing, and feature-fusion models \cite{Yang2024ChannelEstimation,ZhangFengXuLuZhou2018MultiReceiver,CuiZhengLiLiJiang2026WCCNet}. These efforts indicate that robustness is not merely a matter of increasing backbone capacity, but also depends on how models represent, compensate, and fuse signal evidence under changing conditions.

Nevertheless, simulation-to-real generalization remains underexplored. Existing evaluations often rely on transfer learning with measured data from the target scenario, training and testing within a single sea trial, or testing under one limited simulated mismatch. Such settings are useful for feasibility studies, but they do not fully reflect non-cooperative deployment, where target-scenario sea-trial data are typically unavailable beforehand and multiple shift factors may appear together. A protocol that evaluates controlled simulated shifts and zero-shot simulation-to-sea-trial generalization as distinct test settings is therefore needed to assess whether a method is genuinely robust or merely adapted to a specific condition.

\subsection{Underwater Acoustic Modulation Recognition Datasets}\label{subsec:related_benchmark}
Dataset design determines what kinds of robustness claims can be meaningfully supported. Wireless AMC has benefited from reusable public benchmarks such as DeepSig/RadioML, which enable model comparison under shared label spaces and generation protocols \cite{DeepSigDatasets}. By contrast, underwater acoustic modulation recognition still lacks a widely adopted benchmark with comparable standardization.

As summarized in Table~\ref{tab:comparison}, representative underwater acoustic modulation recognition evaluation settings differ substantially in shift coverage, sea-data usage, dataset scale, and open-source availability. Measured sea-trial evaluations provide realism but are usually confined to a specific collection campaign \cite{Zhang2022RCNN}. Replayed-channel and simulation-based studies improve controllability and enable certain cross-scenario tests, but their protocols remain paper-specific and typically cover only part of the robustness problem \cite{Li2024TSTR,Wang2024FeatureFusion,WangHuangShiMao2024}. Public measured-channel replay resources such as Watermark provide reusable channel realizations for underwater acoustic communication evaluation~\cite{VanWalree2017Watermark}, however, they do not provide standardized labels, splits, or shift-specific evaluation protocols for modulation recognition. These heterogeneous settings make it difficult to separate genuine robustness from adaptation to a particular dataset, channel configuration, or evaluation protocol. Consequently, there is still no widely adopted underwater modulation benchmark that jointly supports fair comparison across distribution shifts. This benchmark gap is one reason why benchmark design and model evaluation need to be considered together in the present work.

\begin{table*}[t]
\caption{Comparison of representative underwater acoustic modulation recognition evaluation settings}
\label{tab:comparison}
\centering
\footnotesize
\renewcommand{\arraystretch}{1.1}
\setlength{\tabcolsep}{3pt}
\begin{tabular}{@{}>{\raggedright\arraybackslash}m{0.19\textwidth}>{\raggedright\arraybackslash}m{0.30\textwidth}*{4}{>{\centering\arraybackslash}m{0.055\textwidth}}>{\centering\arraybackslash}m{0.105\textwidth}>{\centering\arraybackslash}m{0.075\textwidth}@{}}
\toprule
Benchmark & Evaluation setting &
\multicolumn{4}{c}{%
\begin{tabular}{@{}*{4}{>{\centering\arraybackslash}m{0.055\textwidth}}@{}}
\multicolumn{4}{c}{Shift coverage} \\
\cmidrule(lr){1-4}
Env. & \makecell{Low\\SNR} & \makecell{Signal\\param.} & \makecell{Sea\\data}
\end{tabular}} &
Samples & \makecell{Open\\source} \\
\midrule
Zhang et al.~\cite{Zhang2022RCNN}
& Training and testing on fully measured signals collected within the same sea trial
& \xmark & \xmark & \xmark & \cmark & 3.0K & \makecell{Not\\Avail.} \\

Li et al.~\cite{Li2024TSTR}
& Cross-environment generalization under replayed measured underwater channels
& \cmark & \xmark & \xmark & \xmark & 108K & \makecell{Not\\Avail.} \\

Wang et al.~\cite{Wang2024FeatureFusion}
& Cross-scenario simulation evaluation with transfer learning to measured sea-trial data
& \cmark & \xmark & \xmark & \cmark & \makecell{NR\textsuperscript{a}\\(1.2K sea test)} & \makecell{Not\\Avail.} \\

\midrule
\addlinespace[1pt]
\makecell[l]{\textbf{UAMR-ShiftBench}\\\textbf{(ours)}}
& \textbf{Unified benchmark covering three representative shifts and sea-data evaluation}
& \textbf{\cmark} & \textbf{\cmark} & \textbf{\cmark} & \textbf{\cmark} & \makecell{\textbf{171.85K}\\\textbf{(2.1K sea test)}} & \textbf{Avail.} \\
\bottomrule
\multicolumn{8}{@{}l@{}}{\footnotesize \textsuperscript{a} NR indicates that the total dataset size is not reported.}
\end{tabular}
\end{table*}

\section{Problem Formulation and UAMR-ShiftBench}\label{sec:uamr_shiftbench_suite}\label{sec:problem_signal_model}

\subsection{Problem Setup}\label{subsec:problem_formulation}
UAMR-ShiftBench defines underwater acoustic modulation recognition as a closed-set waveform classification task under multiple distribution shifts. Let $\mathbf{r}=[r[0],r[1],\ldots,r[T-1]]^{\top}\in\mathbb{R}^{T}$ denote a received passband waveform segment, and let $y\in\mathcal{Y}$ denote its class label.  The label space is fixed across all training and evaluation domains,
$\mathcal{Y}=\{\mathrm{2FSK},\mathrm{4FSK},\mathrm{BPSK},\mathrm{QPSK},\mathrm{OFDM},\mathrm{CW},\mathrm{LFM}\}$. The first five classes correspond to representative communication waveforms, and CW and LFM are included to cover additional waveform types frequently encountered in passive underwater acoustic scenes.

UAMR-ShiftBench provides a unified waveform-level protocol for evaluating recognition models under distribution shift.  Let $\Phi(\mathbf{r}; f_0,B)$ denote the input representation constructed from the received segment, where $f_0$ and $B$ denote the carrier frequency and occupied bandwidth.  A recognition model then predicts
\begin{equation}
f_{\theta}\!\left(
\Phi(\mathbf{r}; f_{0}, B)
\right)
=
\hat{\mathbf{z}}
\in
\mathbb{R}^{C},
\end{equation}
where $C=|\mathcal{Y}|$, $f_{\theta}(\cdot)$ is the learnable classifier, and $\hat{\mathbf{z}}$ is the predicted logit vector.

Let the training set be
\begin{equation}
\mathcal{D}_{\mathrm{train}}=\left\{\left(\mathbf{r}_i,f_{0,i},B_i,y_i\right)\right\}_{i=1}^{N},
\end{equation}
where $N$ denotes the number of labeled training samples. In the present work, the recognition problem remains closed-set, namely $\mathcal{Y}_{\mathrm{train}}=\mathcal{Y}_{\mathrm{test}}=\mathcal{Y}$, but the training and test samples may follow different conditional input distributions,
\begin{equation}
p_{\mathrm{train}}\!\left(\Phi(\mathbf{r};f_0,B)\mid y\right)
\neq
p_{\mathrm{test}}\!\left(\Phi(\mathbf{r};f_0,B)\mid y\right).
\end{equation}
where the mismatch arises from one or more of the four distribution-shift types considered in this paper: i) \emph{\textbf{SNR shift}}, induced by differences in noise level between training and test conditions; ii) \emph{\textbf{environment shift}}, induced by unseen propagation conditions and channel responses; iii) \emph{\textbf{signal-parameter shift}}, induced by held-out communication parameters, specifically unseen carrier frequencies and symbol rates in this work; and iv) \emph{\textbf{simulation-to-real shift}}, induced by the discrepancy between simulated samples and measured sea-trial data. The task is therefore to recognize waveform classes from received passband observations while remaining robust to changes in noise level, propagation environment, and signal configuration.

\subsection{Benchmark Design}\label{subsec:shift_taxonomy}
Building on the shifted closed-set problem setup above, the design goal of UAMR-ShiftBench is therefore to provide a controlled and fair basis for robustness evaluation with a fixed waveform taxonomy, consistent segment format and preprocessing protocol, and structured shift-specific evaluation subsets. This design allows performance degradation to be attributed to the intended mismatch factors and makes results comparable across models.

Due to the scarcity and high collection cost of measured underwater acoustic modulation data, UAMR-ShiftBench adopts a staged evaluation design: simulated data are used for training, large-scale simulated subsets are used for controlled OOD testing, and independent measured sea-trial data are used for the final zero-shot transfer assessment. The measured sea-trial data were acquired during two independent field campaigns conducted in March and November 2025 at shallow-water sites in the South China Sea. This design supports scalable robustness evaluation while retaining a direct test on real shallow-water recordings.

The benchmark has three key characteristics. 1) \emph{\textbf{Shift-disentangled evaluation}}: low-SNR, unseen-environment, unseen-communication-parameter, and simulation-to-real conditions are organized as separate subsets rather than merged into a single aggregate test set. 2) \emph{\textbf{Broad parameter coverage}}: compared with narrow paper-specific settings, UAMR-ShiftBench covers a wider signal-parameter space, including carrier frequencies from 0.8 to 25 kHz and broad symbol-rate or bandwidth ranges, which makes the held-out communication-parameter test more representative of non-cooperative deployment. 3) \emph{\textbf{Realism}}: the simulated propagation environments are generated from real location- and month-dependent sound-speed profiles, and the benchmark further includes two measured sea-trial subsets for direct simulation-to-real assessment. 

\subsection{Benchmark Construction}\label{subsec:benchmark_construction}\label{subsec:exp_dataset}

The construction of UAMR-ShiftBench starts from a passband waveform generation model and then introduces propagation and noise effects to form controlled benchmark subsets.  Let \(x(t;\boldsymbol{\xi})\) denote the transmitted passband waveform, where \(\boldsymbol{\xi}\) is a waveform-dependent parameter vector.  
The received waveform at a passive receiver is affected by multipath propagation and ambient ocean noise \cite{StojanovicPreisig2009,Huang2024BroadbandPINN,Huang2026ParabolicPINN}.  We represent the dominant distortions as
\begin{equation}
r(t)=\sum_{l=1}^{L} \alpha_l(t)\,x\!\left(t-\tau_l;\boldsymbol{\xi}\right)+w(t),
\label{eq:rx_signal_model}
\end{equation}
where \(L\) is the number of effective propagation paths, \(\alpha_l(t)\) denotes the time-varying attenuation of the \(l\)-th path, \(\tau_l\) is the corresponding propagation delay, and \(w(t)\) represents ambient noise. Variations in \(\alpha_l(t)\), \(\tau_l\), \(\boldsymbol{\xi}\), and \(w(t)\) provide the physical basis for the distribution shifts considered in UAMR-ShiftBench. The present work mainly considers stationary source--receiver geometries; therefore, Doppler-induced frequency shifts and time-scale variations are not explicitly introduced.

\begin{table}[!t]
\caption{Summary of the UAMR-ShiftBench Configuration}
\label{tab:config}
\centering
\scriptsize
\renewcommand{\arraystretch}{1.08}
\begin{tabular}{@{}p{0.31\columnwidth}p{0.63\columnwidth}@{}}
\toprule
\textbf{Item} & \textbf{Configuration} \\
\midrule
\multicolumn{2}{@{}>{\columncolor{TableMetric}[0pt][0pt]}p{\dimexpr0.94\columnwidth+2\tabcolsep\relax}@{}}{\textit{Benchmark specification}} \\
Signal classes & 2FSK, 4FSK, BPSK, QPSK, OFDM, CW, and LFM \\
Waveform format & 128 kHz sampling rate; approximately 3 s signal duration (the proposed method uses a 1 s segment) \\
Frequency coverage & 0.8--25 kHz; eight non-overlapping bands \\
Noise setting & AWGN / measured sea noise = 50\% / 50\% \\
Propagation environments & 12 shallow-water environments from three South China Sea locations and WOA monthly sound-speed profiles: 8 seen / 4 unseen\textsuperscript{a} \\
\midrule
\multicolumn{2}{@{}>{\columncolor{TableMetric}[0pt][0pt]}p{\dimexpr0.94\columnwidth+2\tabcolsep\relax}@{}}{\textit{Training and in-distribution evaluation}} \\
In-distribution set & SNR = [0, 20] dB; 11,250/class; 75\% / 10\% / 15\% train/validation/test split \\
\midrule
\multicolumn{2}{@{}>{\columncolor{TableMetric}[0pt][0pt]}p{\dimexpr0.94\columnwidth+2\tabcolsep\relax}@{}}{\textit{Distribution-shift evaluation}} \\
Low-SNR set & SNR = [-8, 0) dB; 3,000/class \\
Unseen-environment set & SNR = [-8, 20] dB; 4,000/class; 4 unseen propagation environments \\
Unseen-parameter set & SNR = [-8, 20] dB; 6,000/class; held-out frequency bands and symbol-rate or bandwidth settings \\
Sea-trial set & Two South China Sea shallow-water sites;\newline March: 100/class; November: 200/class \\
\bottomrule
\end{tabular}
\vspace{0.6mm}
\parbox{0.94\columnwidth}{\footnotesize \textsuperscript{a} Auxiliary NoChannel samples are used for training diversity.}
\vspace{-2 em}
\end{table}
For recognition, the continuous-time waveform in Equation~\eqref{eq:rx_signal_model} is sampled over one observation interval to form the discrete segment $\mathbf{r}$ used in the problem setup. Table~\ref{tab:config} summarizes the main benchmark configuration, including the waveform classes, frequency coverage, propagation environments, noise setting, subset sizes, and split protocol. Detailed signal-parameter ranges and construction rules are provided in~\ref{app:dataset_config}. Because the waveform classes occupy different bandwidths, SNR is defined as the in-band signal-to-noise ratio,
\begin{equation}
\mathrm{SNR}_{\mathrm{in\mbox{-}band}} = 10 \log_{10}\!\left(P_s^{(\mathcal{B})}/P_n^{(\mathcal{B})}\right)\ \mathrm{dB},
\end{equation}
where $P_s^{(\mathcal{B})}$ and $P_n^{(\mathcal{B})}$ denote the signal and noise power measured within the signal bandwidth $\mathcal{B}$, respectively. 

The simulated environments are built around shallow-water South China Sea locations to match the geographic regime of the measured sea-trial data while still allowing controlled environment separation. Figure~\ref{fig:benchmark_environment} shows the three simulation locations, the measured sea-trial areas, and the sound-speed profiles used in UAMR-ShiftBench. Locations L1 and L2 are close to the two sea-trial areas and lie on the continental shelf, whereas L3 is placed farther from L1/L2 and on the continental-slope region to create a more distinct propagation environment. WOA profiles are used because they provide reproducible location- and month-dependent sound-speed fields, enabling seasonal channel variability to be introduced without site-specific manual tuning \cite{WOA}. Holding out L3 for testing creates a spatial environment shift, while the March and November sea-trial profiles provide the final simulation-to-real assessment, as shown in Fig.~\ref{fig:benchmark_environment}(b) and (c).

\begin{figure*}[t]
    \centering
    \includegraphics[width=1\textwidth]{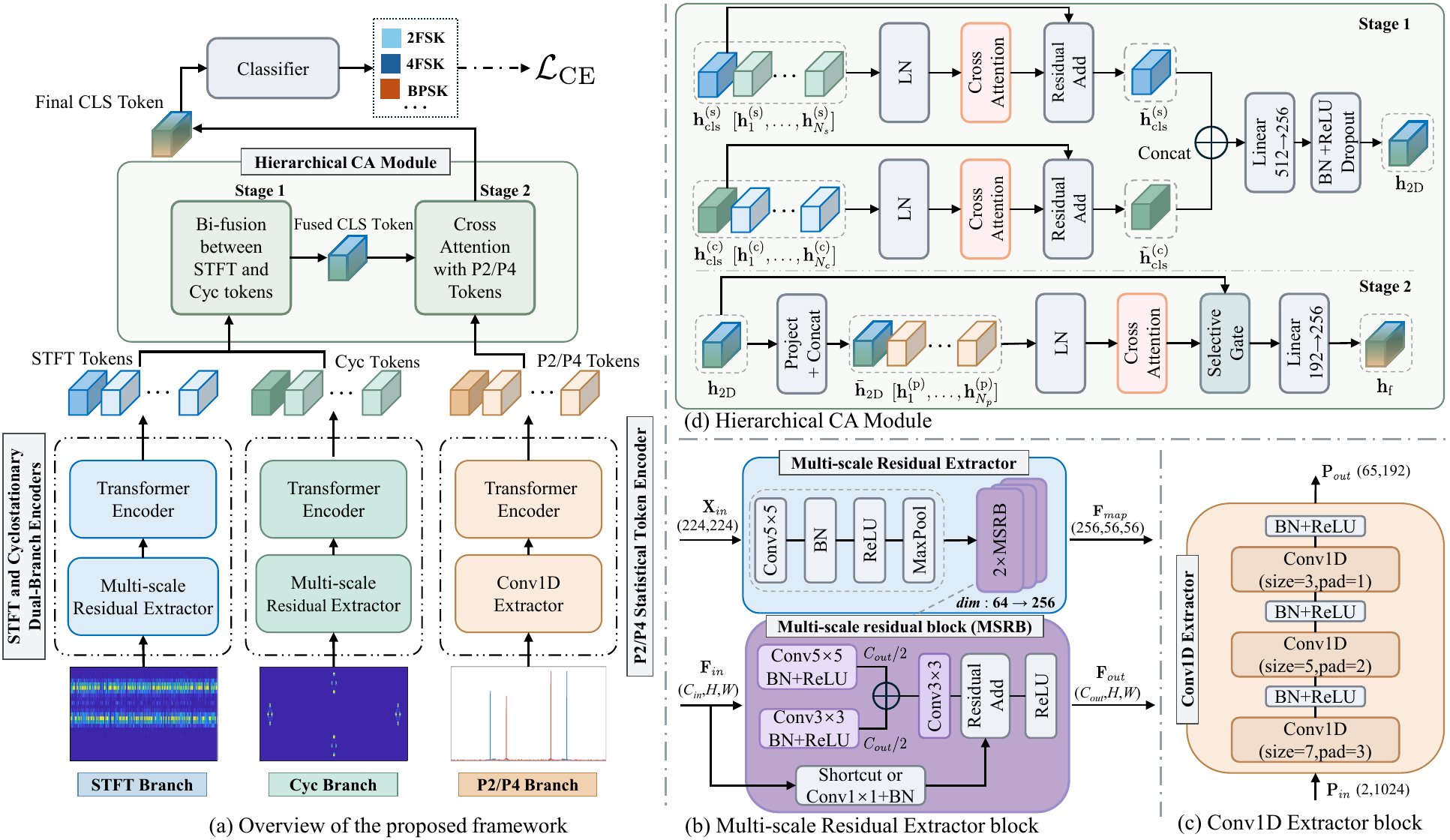}
    \caption{Overall architecture of SCP-TriCA. 
    (a) Overview of the proposed framework. 
    (b) Multi-scale residual extractor. 
    (c) Conv1D extractor. 
    (d) Hierarchical cross-attention fusion module.
    }
    \label{fig:scptrica_overall}
\end{figure*}
\section{Methodology}\label{sec:method}

\subsection{Overall Architecture of SCP-TriCA}\label{subsec:method_overall}

SCP-TriCA processes each received waveform segment through three parallel branches and a hierarchical fusion stage, as illustrated in Fig.~\ref{fig:scptrica_overall}(a). The three branches handle complementary signal representations: the STFT branch captures time-frequency energy structure, the cyclostationary branch emphasizes modulation-induced periodic correlations, and the P2/P4 branch encodes higher-order spectral statistics. Each branch consists of a dedicated feature extractor followed by a transformer encoder that produces a sequence of tokens together with a global cls token summarizing the branch-level evidence. Fusion proceeds in two stages. In Stage 1, the STFT and cyclostationary cls tokens exchange information through bidirectional cls-guided cross-attention, and the two updated tokens are compressed into a unified 2D representation. In Stage 2, this fused representation queries the P2/P4 token sequence through a second cross-attention block, and a sample-adaptive selective gate controls how much statistical evidence is incorporated before final classification.
The rationale for this two-stage hierarchy is as follows. The STFT and cyclostationary modalities share the same 2D token geometry and jointly encode the dominant discriminative cues under most conditions, so they are fused first. The P2/P4 statistical descriptors provide complementary higher-order cues that are particularly informative for hard-to-separate classes such as BPSK and QPSK, and for simulation-to-real transfer, but are more sensitive to severe noise mismatch; injecting them as a controlled refinement after the 2D fusion therefore improves robustness without risking early corruption of the shared representation. The three input modalities fed into this architecture are constructed as described in Section~\ref{subsec:method_inputs}.

\subsection{Tri-Modal Input Construction}\label{subsec:method_inputs}

\begin{figure*}[t]
\centering
\includegraphics[width=0.92\textwidth]{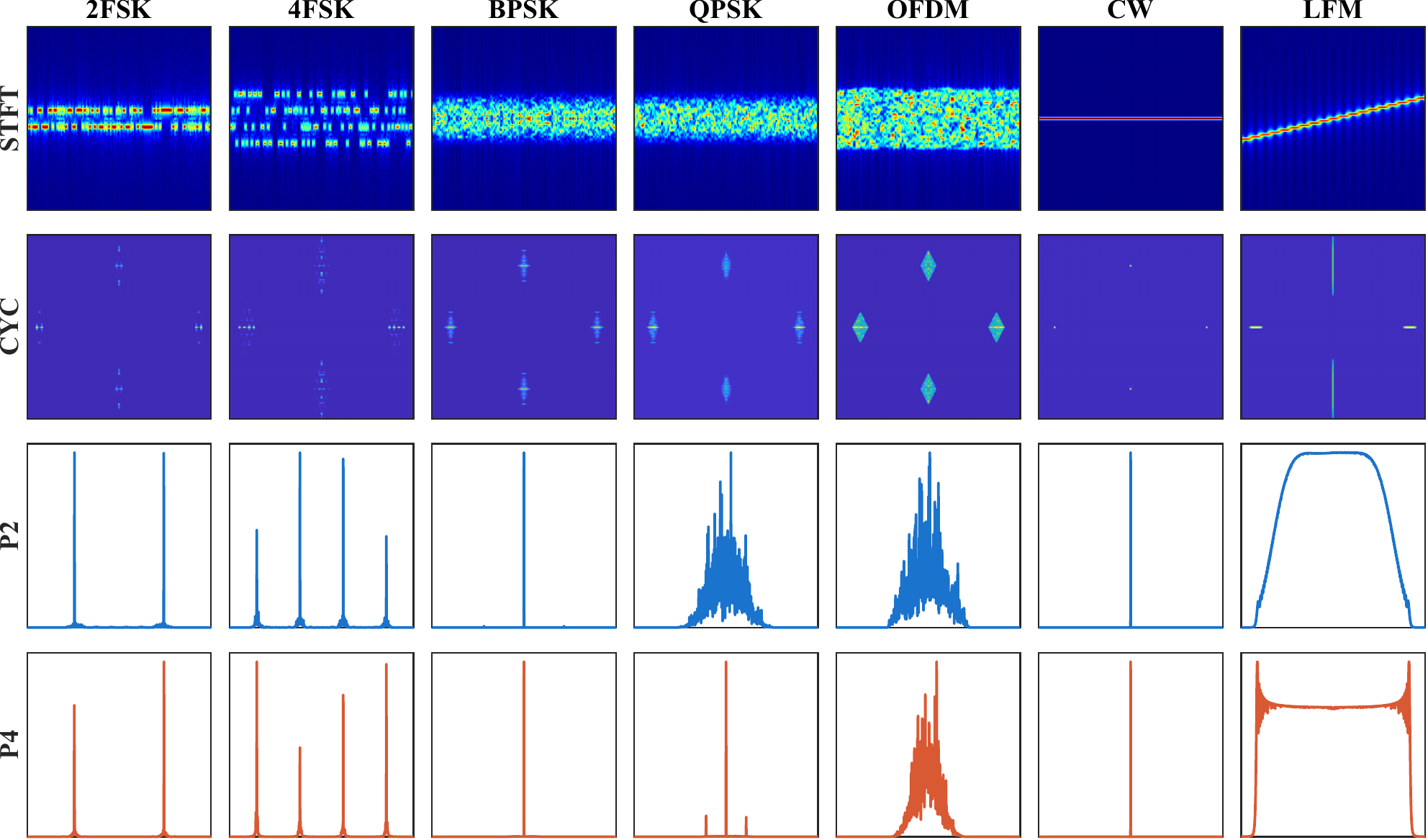}%
\caption{Tri-modal examples of the seven pure-signal classes at a carrier frequency of 7 kHz, showing the STFT, cyclostationary, and P2/P4 modalities used in the proposed preprocessing pipeline.}
\label{fig:app_seven_signals_trimodal_fc7k}
\end{figure*}

For SCP-TriCA, the representation $\Phi(\mathbf{r}; f_{\mathrm{L}},f_{\mathrm{H}})$ is instantiated as a tri-modal tuple for each 1 s received segment $\mathbf{r}=[r[0],\ldots,r[T-1]]^{\top}$, where $T=128{,}000$ at the 128 kHz sampling rate:
\begin{equation}
\left(\mathbf{X}^{\mathrm{stft}},\mathbf{X}^{\mathrm{cyc}},(\mathbf{p}_2,\mathbf{p}_4)\right)
= \Phi(\mathbf{r};f_{\mathrm{L}},f_{\mathrm{H}}),
\end{equation}
where $\mathbf{X}^{\mathrm{stft}}\in[0,1]^{224\times224}$ denotes the normalized time-frequency matrix, $\mathbf{X}^{\mathrm{cyc}}\in[0,1]^{224\times224}$ denotes the normalized cycle frequency-frequency matrix, and $\mathbf{p}_2,\mathbf{p}_4\in[0,1]^{1024}$ denote the normalized second- and fourth-order power-spectrum descriptors. Each modality is derived from the same received segment using the carrier-frequency and bandwidth information $(f_0,B)$.
These modalities provide complementary time-frequency, cyclostationary, and higher-order statistical views, motivating the hierarchical fusion strategy in Section~\ref{subsec:method_hierarchical_fusion}.

The concrete construction of each modality is detailed below. For the STFT modality, the occupied subband is selected from the spectrogram and converted into a normalized time-frequency matrix,
\begin{equation}
\mathbf{X}^{\mathrm{stft}}
= \mathcal{N}\!\left(
\mathcal{R}_{224}\!\left(
\left.P_{\mathrm{stft}}(t,f)\right|_{(t,f)\in\Omega_{\mathrm{stft}}}
\right)\right),
\end{equation}
where $P_{\mathrm{stft}}(t,f)$ denotes the normalized STFT log-power spectrum feature, $\Omega_{\mathrm{stft}}$ denotes the retained subband display region, $\mathcal{N}(\cdot)$ denotes min-max normalization, and $\mathcal{R}_{224}(\cdot)$ denotes resizing to $224\times224$. 

For the cyclostationary modality, the received signal is first bandpass-filtered according to the occupied band, after which its spectral-correlation map is estimated by the averaged cyclic periodogram (ACP) method \cite{Antoni2007CyclicSpectralPractice},
\begin{equation}
\mathbf{X}^{\mathrm{cyc}}
= \mathcal{N}\!\left(
\mathcal{R}_{224}\!\left(
\left.S(\alpha,f)\right|_{(\alpha,f)\in\Omega_{\mathrm{cyc}}}
\right)\right),
\end{equation}
where $S(\alpha,f)$ denotes the log-cyclic spectrum and $\Omega_{\mathrm{cyc}}$ denotes the retained display region. 

For the P2/P4 modality, the normalized second- and fourth-order descriptors are formed as
\begin{equation}
\begin{aligned}
\mathbf{p}_M
&= \mathcal{H}_{1024}\!\left(
\mathcal{N}\!\left(
\left.P_M[\ell]\right|_{\mathcal{B}_M}
\right)\right),\\
\mathcal{B}_M
&=\{\,\ell \mid Mf_{\mathrm{L}}\leq \ell\Delta f_M \leq Mf_{\mathrm{H}}\,\},
\; M\in\{2,4\},
\end{aligned}
\end{equation}
where $P_M[\ell]$ denotes the $M$th-order power spectrum and $\mathcal{H}_{1024}(\cdot)$ denotes fixed-length resampling. 
Representative examples of the resulting tri-modal inputs are shown in Fig.~\ref{fig:app_seven_signals_trimodal_fc7k}.

\subsection{STFT and Cyclostationary Dual-Branch Encoders}\label{subsec:method_dual_branch}
Given the STFT matrix $\mathbf{X}^{\mathrm{stft}}$ and cyclostationary matrix $\mathbf{X}^{\mathrm{cyc}}$ constructed in Section~\ref{subsec:method_inputs}, the two 2D branches encode each input into a sequence of local patch tokens together with a global cls token. The two branches share the same encoder architecture but maintain independent parameters, keeping them geometrically aligned for later cross-modal interaction while allowing each to adapt to its own input modality. As shown in Fig.~\ref{fig:scptrica_overall}(a) and Fig.~\ref{fig:scptrica_overall}(b), for each modality $m\in\{\mathrm{stft},\mathrm{cyc}\}$, the front-end first applies one convolutional stem and three residual stages,
\begin{equation}
\begin{aligned}
\mathbf{F}_{0}^{(m)}
&=\operatorname{MP}\!\left(\operatorname{ReLU}\!\left(\operatorname{BN}\!\left(\operatorname{Conv}_{5\times5}^{(m)}(\mathbf{X}^{(m)})\right)\right)\right),\\
\mathbf{F}_{l}^{(m)}
&=\mathcal{R}_{l}^{(m)}\!\left(\mathbf{F}_{l-1}^{(m)}\right),\quad l=1,2,3,
\end{aligned}
\end{equation}
where $\operatorname{Conv}_{5\times5}^{(m)}(\cdot)$ denotes the initial $5\times5$ convolution, $\operatorname{MP}(\cdot)$ denotes max-pooling, and $\mathcal{R}_{l}^{(m)}(\cdot)$ denotes the $l$th multi-scale residual stage of modality $m$. The three residual stages progressively increase the channel width as $64\!\rightarrow\!64\!\rightarrow\!128\!\rightarrow\!256$.

Each residual stage is built from a multi-scale residual block. At the $l$th stage of modality $m$, given the input feature map $\mathbf{F}_{l-1}^{(m)}$, the block produces $\mathbf{F}_{l}^{(m)}$ by computing
\begin{equation}
\begin{aligned}
\mathbf{U}_{3,l}^{(m)}
&=\operatorname{ReLU}\!\left(\operatorname{BN}\!\left(
\operatorname{Conv}_{3\times3}(\mathbf{F}_{l-1}^{(m)})
\right)\right),\\
\mathbf{U}_{5,l}^{(m)}
&=\operatorname{ReLU}\!\left(\operatorname{BN}\!\left(
\operatorname{Conv}_{5\times5}(\mathbf{F}_{l-1}^{(m)})
\right)\right),\\
\mathbf{F}_{l}^{(m)}
&=\operatorname{ReLU}\!\Big(
\operatorname{BN}\!\left(
\operatorname{Conv}_{3\times3}^{\mathrm{post}}\!\left(
[\mathbf{U}_{3,l}^{(m)};\mathbf{U}_{5,l}^{(m)}]
\right)\right)\\
&\quad+\operatorname{Proj}_{s}(\mathbf{F}_{l-1}^{(m)})
\Big),
\qquad l=1,2,3.
\end{aligned}
\end{equation}
where $\operatorname{Conv}_{3\times3}(\cdot)$ and $\operatorname{Conv}_{5\times5}(\cdot)$ denote the parallel $3\times3$ and $5\times5$ convolutions, $\operatorname{Conv}_{3\times3}^{\mathrm{post}}(\cdot)$ denotes the post-fusion $3\times3$ convolution, and $\operatorname{Proj}_{s}(\cdot)$ denotes the shortcut path implemented as an identity mapping or a $1\times1$ projection when dimension matching is required. The parallel $3\times3$ and $5\times5$ paths capture local and broader spectral patterns, following multi-scale designs used in recent UAMR networks~\cite{Huang2026Hybrid,WangHuangShiMao2024,Wang2024FeatureFusion}.
The final 2D feature maps are then converted into fixed-cardinality token sequences by adaptive pooling and transformer encoders. For each paired branch-modality index $(m,b)\in\{(\mathrm{stft},\mathrm{s}),(\mathrm{cyc},\mathrm{c})\}$, the pooled token grid is
\begin{equation}
\begin{aligned}
\mathbf{T}^{(b)}
&=\operatorname{Flat}\!\left(
\operatorname{AAP}_{8\times8}\!\left(\mathbf{F}_{3}^{(m)}\right)
\right),\\
\mathbf{T}^{(b)}
&\in\mathbb{R}^{N_{b}\times d},\; N_s=N_c=64, \;d=256,
\end{aligned}
\end{equation}
The grid is then further refined by a one-layer transformer encoder with four attention heads, a learnable class token, and positional embeddings, yielding
\begin{equation}
\left[\mathbf{h}_{\mathrm{cls}}^{(b)},\mathbf{h}_1^{(b)},\ldots,\mathbf{h}_{N_b}^{(b)}\right]
\!=\mathcal{E}_{\mathrm{2D}}^{(m)}\!\left(\mathbf{T}^{(b)}\right)
\in\mathbb{R}^{(1+N_b)\times d}.
\end{equation}
Here $\mathbf{F}_{3}^{(m)}$ denotes the final feature map of modality $m$, $\operatorname{AAP}_{8\times8}(\cdot)$ produces an $8\times8$ pooled grid, $\operatorname{Flat}(\cdot)$ reshapes this grid into $N_b=64$ local tokens, and $\mathcal{E}_{\mathrm{2D}}^{(m)}(\cdot)$ denotes the corresponding one-layer transformer encoder. The output token $\mathbf{h}_{\mathrm{cls}}^{(b)}$ summarizes global branch-level evidence, while $\mathbf{h}_{i}^{(b)}$ denotes the $i$th local patch token. These STFT and cyclostationary tokens are then fed into the hierarchical CA module described in Section~\ref{subsec:method_hierarchical_fusion}.

\subsection{P2/P4 Statistical Token Encoder}\label{subsec:method_p2p4_branch}
Unlike the two 2D modalities, the P2/P4 descriptors $\mathbf{p}_2$ and $\mathbf{p}_4$ constructed in Section~\ref{subsec:method_inputs} are one-dimensional statistical sequences rather than spatial maps. Therefore, SCP-TriCA uses a dedicated 1D token encoder for this branch. As shown in Fig.~\ref{fig:scptrica_overall}(a) and Fig.~\ref{fig:scptrica_overall}(c), the encoder stacks $\mathbf{p}_2$ and $\mathbf{p}_4$ into a two-channel sequence $[\mathbf{p}_2;\mathbf{p}_4]\in\mathbb{R}^{2\times1024}$ and applies three Conv1D--BN--ReLU blocks with channel progression $2\!\rightarrow\!64\!\rightarrow\!128\!\rightarrow\!192$ and kernel sizes $7$, $5$, and $3$, respectively. These convolutional layers extract local spectral patterns and progressively mix the P2/P4 channels before tokenization.

The resulting 1D feature map is then compressed by adaptive average pooling into $N_p=64$ bins and further refined by a one-layer transformer encoder with four attention heads, a learnable cls token, and positional embeddings, yielding
\begin{equation}
\begin{aligned}
&\left[\mathbf{h}_{\mathrm{cls}}^{(\mathrm{p})},
\mathbf{h}_1^{(\mathrm{p})},\ldots,
\mathbf{h}_{N_p}^{(\mathrm{p})}\right]\\
&\quad=\mathcal{E}_{\mathrm{1D}}\!\left(
\operatorname{AAP}_{N_p}\!\left(\mathbf{F}^{(\mathrm{p})}\right)
\right)
\in\mathbb{R}^{(1+N_p)\times d_p},
\end{aligned}
\end{equation}
where $\mathbf{F}^{(\mathrm{p})}$ denotes the Conv1D feature map, $\operatorname{AAP}_{N_p}(\cdot)$ produces $N_p=64$ pooled bins, $d_p=192$ is the token dimension, and $\mathcal{E}_{\mathrm{1D}}(\cdot)$ denotes the corresponding one-layer transformer encoder. The output token $\mathbf{h}_{\mathrm{cls}}^{(\mathrm{p})}$ summarizes global P2/P4 statistical evidence, while $\mathbf{h}_{i}^{(\mathrm{p})}$ denotes the $i$th local statistical token. These statistical tokens, together with the STFT and cyclostationary tokens from Section~\ref{subsec:method_dual_branch}, are jointly processed by the hierarchical CA module described next.

\begin{figure}[h]
    \centering
    \includegraphics[width=0.98\linewidth]{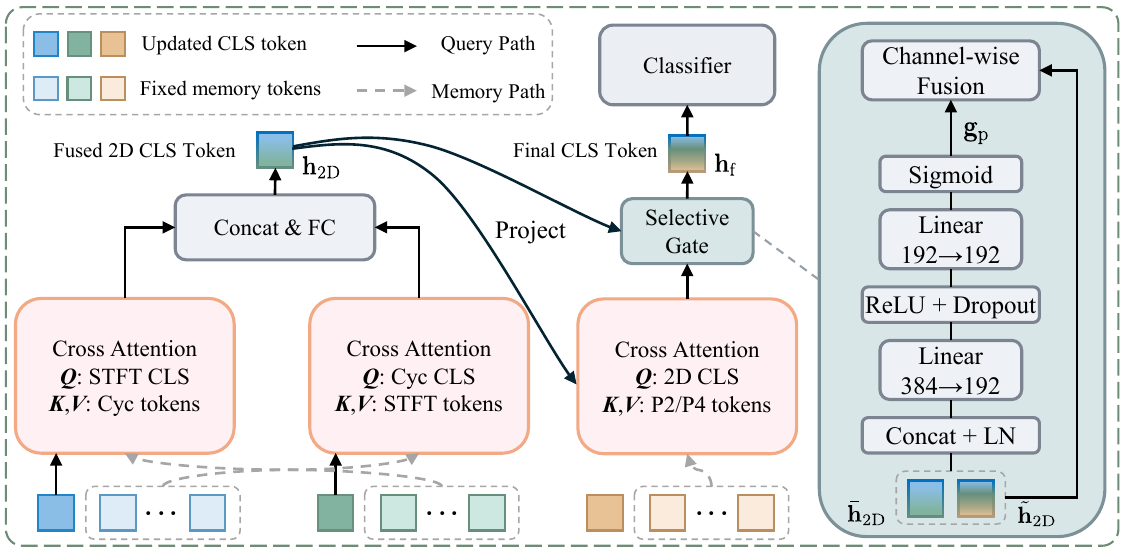}
    \caption{The hierarchical fusion mechanism in SCP-TriCA.}
    \label{fig:hierarchical_fusion_mechanism}
\end{figure}
\vspace{-1.45em}

\subsection{Hierarchical Fusion Mechanism}\label{subsec:method_hierarchical_fusion}
Given the encoded STFT, cyclostationary, and P2/P4 tokens, the hierarchical CA module implements the two-stage fusion outlined in Section~\ref{subsec:method_overall}: bidirectional interaction between the two 2D modalities followed by gated statistical refinement from the P2/P4 branch. Fig.~\ref{fig:scptrica_overall}(d) shows the module structure, and Fig.~\ref{fig:hierarchical_fusion_mechanism} illustrates the corresponding query--memory interaction path. 

The key operator in both stages is a cls-guided cross-attention module \cite{Chen2021CrossViT,Chowdhury2025T3Time,Zhang2025MST}. It is denoted by $\operatorname{CA}(\mathbf{q},\mathbf{M})$, where $\mathbf{q}$ is a single global query token and $\mathbf{M}$ denotes the source-modality memory tokens. Instead of performing full token-to-token interaction, SCP-TriCA updates only this global query while treating the source tokens as an external memory. This restriction keeps the fusion path lightweight, preserves local modality-specific structure, and encourages selective evidence borrowing rather than unrestricted feature mixing. From a robustness perspective, this is desirable because a degraded modality can still provide complementary context without globally overwriting the local encoding of the other branch.

Stage 1 performs bidirectional interaction between the STFT and cyclostationary branches. The two updated global tokens are
\begin{equation}
\begin{aligned}
\tilde{\mathbf{h}}_{\mathrm{cls}}^{(\mathrm{s})}
&=\operatorname{CA}\!\left(
\mathbf{h}_{\mathrm{cls}}^{(\mathrm{s})},
[\mathbf{h}_1^{(\mathrm{c})},\ldots,\mathbf{h}_{N_c}^{(\mathrm{c})}]
\right),\\
\tilde{\mathbf{h}}_{\mathrm{cls}}^{(\mathrm{c})}
&=\operatorname{CA}\!\left(
\mathbf{h}_{\mathrm{cls}}^{(\mathrm{c})},
[\mathbf{h}_1^{(\mathrm{s})},\ldots,\mathbf{h}_{N_s}^{(\mathrm{s})}]
\right),
\end{aligned}
\end{equation}
These two updated global tokens are then concatenated and compressed into a unified 2D representation:
\begin{equation}
\mathbf{h}_{\mathrm{2D}}
=\mathcal{F}_{\mathrm{2D}}\!\left(
\left[\tilde{\mathbf{h}}_{\mathrm{cls}}^{(\mathrm{s})};\tilde{\mathbf{h}}_{\mathrm{cls}}^{(\mathrm{c})}\right]
\right)
\in\mathbb{R}^{256}.
\end{equation}
where $\mathcal{F}_{\mathrm{2D}}(\cdot)$ denotes a learnable mapping from the concatenated global tokens to a 256-dimensional 2D representation. This first stage mainly addresses perturbations that manifest in 2D appearance. Through bidirectional cross-attention, each 2D branch reads complementary global information from the other before final compression.

\begin{table*}[t]
\caption{Overall comparison of \textbf{SCP-TriCA} and representative baselines on \textbf{UAMR-ShiftBench}. The best results are highlighted in \textbf{bold}, and the second-best results are \underline{underscored}.}
\label{tab:overall_comparison}
\centering
\small
\renewcommand{\arraystretch}{1.3}
\setlength{\tabcolsep}{4pt}
\resizebox{\textwidth}{!}{%
\begin{tabular}{lccccccccccc}
\specialrule{1.0pt}{0pt}{0pt}
\specialrule{0.5pt}{0pt}{2pt}
Method
& \begin{tabular}[c]{@{}c@{}}\tablehead{In-Distribution (\%)}\\\cmidrule(lr){1-1}\tablehead{[0, 20] dB}\end{tabular}
& \begin{tabular}[c]{@{}c@{}}\tablehead{Low-SNR (\%)}\\\cmidrule(lr){1-1}\tablehead{[-8, 0) dB}\end{tabular}
& \multicolumn{3}{c}{\begin{tabular}[c]{@{}ccc@{}}\multicolumn{3}{c}{\tablehead{Unseen Environment (\%)}}\\\cmidrule(lr){1-3}\tablehead{[-8, 0) dB} & \tablehead{[0, 20] dB} & \tablehead{Overall}\end{tabular}}
& \multicolumn{3}{c}{\begin{tabular}[c]{@{}ccc@{}}\multicolumn{3}{c}{\tablehead{Unseen Comm. Parameters (\%)}}\\\cmidrule(lr){1-3}\tablehead{[-8, 0) dB} & \tablehead{[0, 20] dB} & \tablehead{Overall}\end{tabular}}
& \tablehead{OOD Avg.\\(\%)}
& \multicolumn{2}{c}{\begin{tabular}[c]{@{}cc@{}}\multicolumn{2}{c}{\tablehead{Sea-Trial (\%)}}\\\cmidrule(lr){1-2}\tablehead{March} & \tablehead{November}\end{tabular}} \\
\midrule
R\&CNN~\cite{Zhang2022RCNN} & 16.61 & 14.90 & 14.01 & 16.49 & 15.25 & 14.49 & 14.37 & 14.43 & 14.86 & 13.00 & 14.29 \\
TRN~\cite{Cai2022TRN} & 51.91 & 34.19 & 33.69 & 53.04 & 43.36 & 31.04 & 49.43 & 40.23 & 39.26 & 29.43 & 43.14 \\
TSTR~\cite{Li2024TSTR} & 79.17& 53.97 & 53.68 & 78.47 & 66.07 & 54.04 & 77.92 & 65.98& 62.01 & 60.71 & \underline{71.86} \\
S\&SEFM~\cite{Wang2024FeatureFusion} & 85.71 & 56.34 & 56.55 & 84.70 & 70.63 & 55.80 & 81.44 & 68.62 & 65.20 & \underline{75.43} & 65.50 \\
IQFormer~\cite{Shao2025IQFormer} & \underline{93.96} & \underline{60.26} & \underline{58.20} & \underline{91.12} & \underline{74.66} & \underline{56.27} & \underline{90.71} & \underline{73.49} & \underline{69.47} & 71.00 & 63.79 \\
\midrule
\rowcolor{TableOurs}
SCP-TriCA (ours) & \textbf{95.33} & \textbf{65.48} & \textbf{64.41} & \textbf{93.70} & \textbf{79.06} & \textbf{64.87} & \textbf{93.58} & \textbf{79.23} & \textbf{74.59} & \textbf{91.14} & \textbf{94.86} \\
\specialrule{0.5pt}{2pt}{0pt}
\specialrule{1.0pt}{0pt}{0pt}
\end{tabular}}
\end{table*}

\begin{figure*}[!t]
\centering
\subfloat[]{%
\includegraphics[width=0.235\textwidth,trim=8 6 8 6,clip]{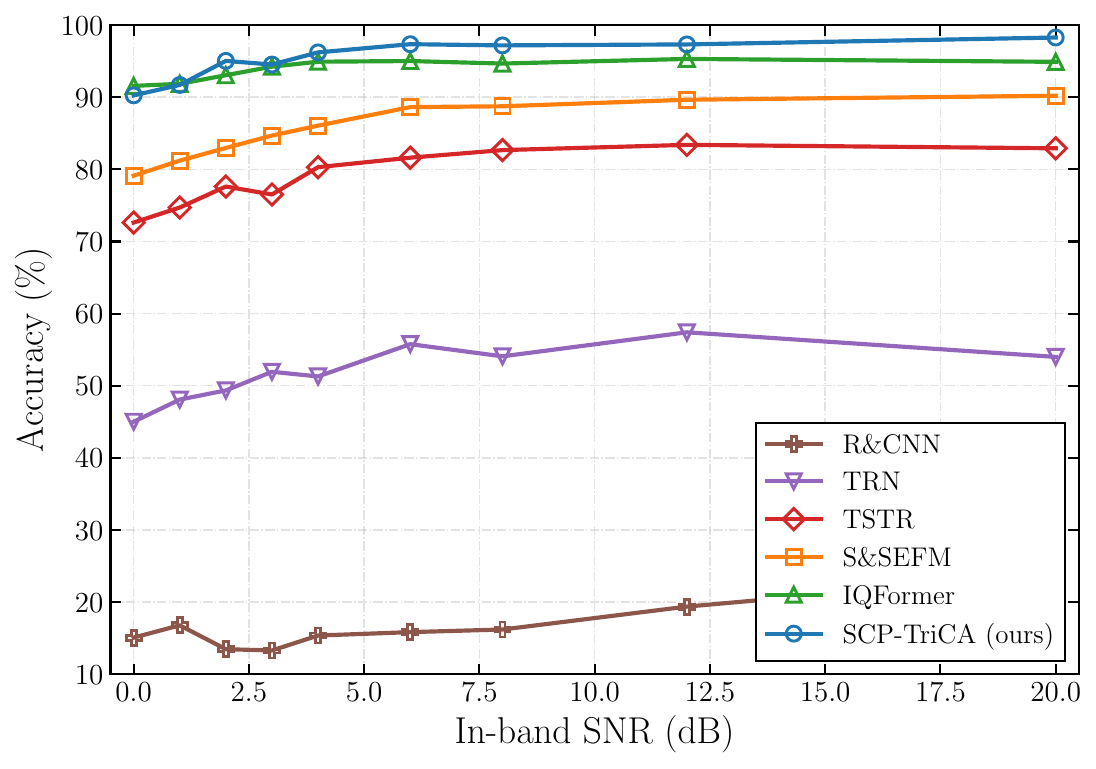}}
\hfil
\subfloat[]{%
\includegraphics[width=0.235\textwidth,trim=8 6 8 6,clip]{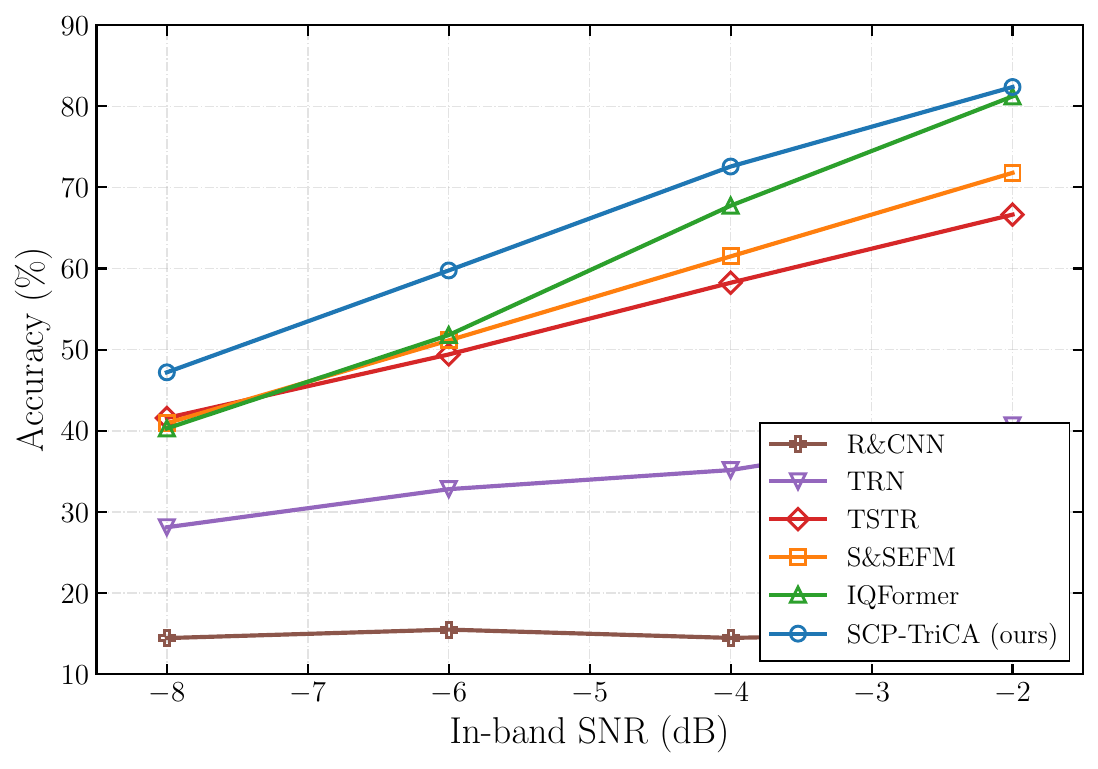}}
\hfil
\subfloat[]{%
\includegraphics[width=0.235\textwidth,trim=8 6 8 6,clip]{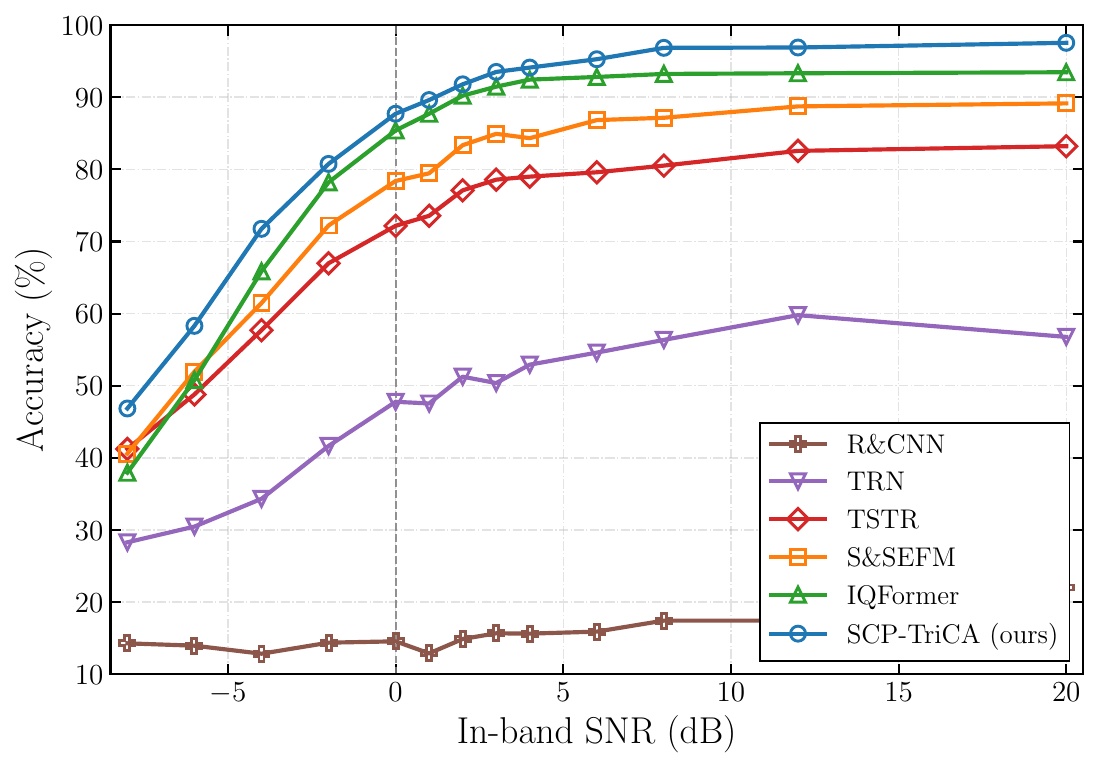}}
\hfil
\subfloat[]{%
\includegraphics[width=0.235\textwidth,trim=8 6 8 6,clip]{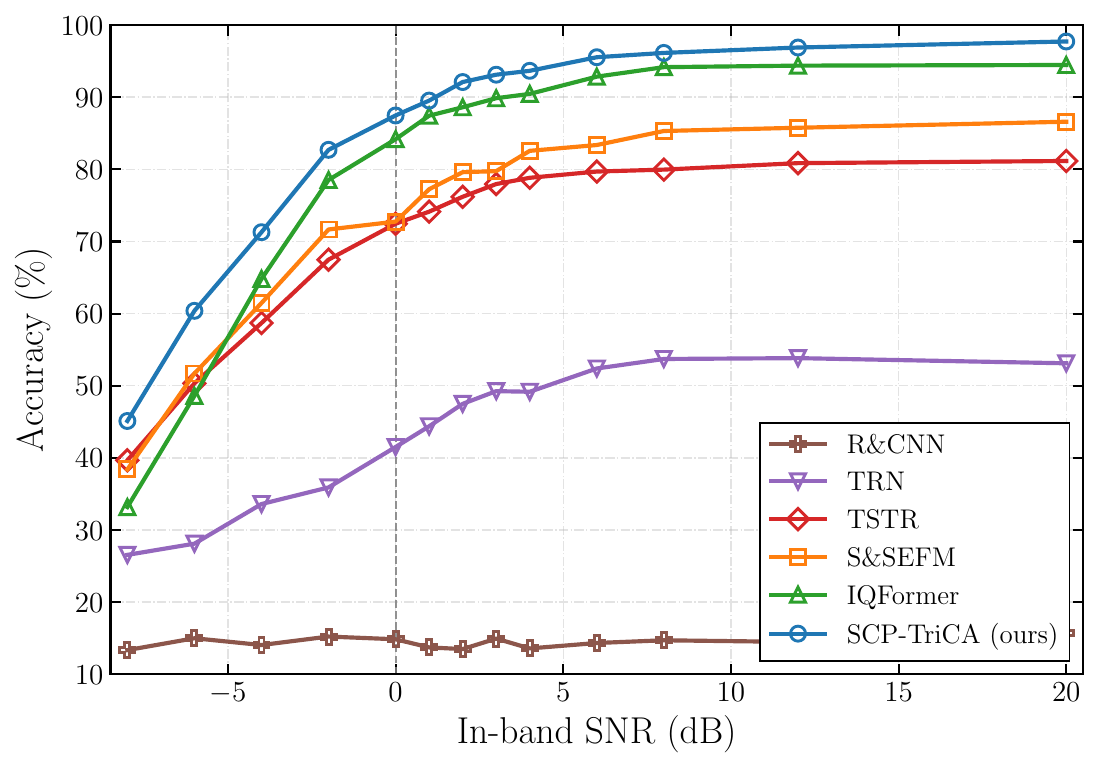}}
\caption{Recognition accuracy comparison of model performance across the four benchmark settings. (a) In-distribution. (b) Low-SNR. (c) Unseen environment. (d) Unseen communication parameter.}
\label{fig:models_compare_d1_d4}
\end{figure*}

Stage 2 uses the fused 2D representation as a query to read the P2/P4 statistical tokens. Because the statistical branch uses a different token dimension, the fused 2D token is first projected into the P2/P4 space and then updated through cls-guided cross-attention:
\begin{equation}
\begin{aligned}
\bar{\mathbf{h}}_{\mathrm{2D}}
&=\mathcal{P}_{\mathrm{2D}\rightarrow \mathrm{p}}\!\left(
\mathbf{h}_{\mathrm{2D}}
\right),\\
\tilde{\mathbf{h}}_{\mathrm{2D}}
&=\operatorname{CA}\!\left(
\bar{\mathbf{h}}_{\mathrm{2D}},
[\mathbf{h}_1^{(\mathrm{p})},\ldots,\mathbf{h}_{N_p}^{(\mathrm{p})}]
\right).
\end{aligned}
\end{equation}
Here $\mathcal{P}_{\mathrm{2D}\rightarrow \mathrm{p}}(\cdot)$ denotes a learnable linear projection from the 2D fusion space to the P2/P4 token space.
A fixed-weight injection would assume that the P2/P4 statistical evidence is equally reliable for all samples, which is unlikely under sea-trial conditions where higher-order statistics can be affected differently by ambient noise, multipath propagation, and channel variability~\cite{StojanovicPreisig2009,Antoni2007CyclicSpectralPractice}. To make the statistical refinement conditional on the current sample, the model applies a sample-adaptive channel-wise gate that interpolates between the original projected query and the P2/P4-updated query:

\begin{equation}
\mathbf{g}_{\mathrm{p}}
=\operatorname{Sigmoid}\!\left(
\mathcal{G}_{\mathrm{p}}\!\left(
[\bar{\mathbf{h}}_{\mathrm{2D}};\tilde{\mathbf{h}}_{\mathrm{2D}}]
\right)\right),
\end{equation}
where $\mathcal{G}_{\mathrm{p}}(\cdot)$ is a two-layer gating MLP consisting of LayerNorm, Linear$(2d_p,d_p)$, GELU, dropout, and Linear$(d_p,d_p)$, followed by sigmoid activation. Thus, $\mathbf{g}_{\mathrm{p}}\in[0,1]^{d_p}$ controls the per-sample, per-channel contribution of the P2/P4-updated query. The final Stage-2 representation is obtained by channel-wise fusion:
\begin{equation}
\begin{aligned}
\mathbf{h}_{\mathrm{2D}}^{*}
&=(1-\mathbf{g}_{\mathrm{p}})\odot\bar{\mathbf{h}}_{\mathrm{2D}}
+\mathbf{g}_{\mathrm{p}}\odot\tilde{\mathbf{h}}_{\mathrm{2D}},\\
\mathbf{h}_{\mathrm{f}}
&=\mathcal{P}_{\mathrm{p}\rightarrow \mathrm{2D}}\!\left(
\mathbf{h}_{\mathrm{2D}}^{*}
\right)
\in\mathbb{R}^{256}.
\end{aligned}
\end{equation}
Here $\mathcal{P}_{\mathrm{p}\rightarrow \mathrm{2D}}(\cdot)$ projects the gated representation back to the 2D fusion space. Gate values close to zero preserve the projected 2D query, whereas values close to one emphasize the P2/P4-updated query.
The fused representation $\mathbf{h}_{\mathrm{f}}$ is then passed to a lightweight classifier $g(\cdot)$ to produce the logit vector $\hat{\mathbf{z}} = g(\mathbf{h}_{\mathrm{f}})\in\mathbb{R}^{C}$. The model is optimized end-to-end with the standard cross-entropy loss.

\section{Experiments}\label{sec:experiments_results}
\subsection{Compared Methods and Experimental Protocol}\label{subsec:exp_tasks}
We compare SCP-TriCA with five representative baselines: \emph{R}\&\emph{CNN} \cite{Zhang2022RCNN}, \emph{TRN} \cite{Cai2022TRN}, \emph{TSTR} \cite{Li2024TSTR}, \emph{S}\&\emph{SEFM} \cite{Wang2024FeatureFusion}, and \emph{IQFormer} \cite{Shao2025IQFormer}, covering recurrent-convolutional, transformer-based, dual-stream fusion, lightweight feature-fusion, and multimodal recognition designs. All baselines use the same train\slash validation\slash test split, benchmark-defined preprocessing, and optimization protocol. I/Q-based baselines obtain baseband sequences by down-conversion with the carrier frequency $f_0$; for methods requiring different front-end representations, we generate their required inputs from the same received waveform segments.

All models are implemented in PyTorch and trained on a workstation with an NVIDIA GeForce RTX 5090 GPU. We use cross-entropy loss and AdamW with an initial learning rate of $3\times10^{-4}$, weight decay of $10^{-2}$, batch size of $64$, and a maximum of $80$ epochs. The learning rate follows a cosine schedule with $3$ warm-up epochs and a minimum value of $10^{-6}$. Gradients are clipped to a maximum norm of $5.0$, and training stops when validation performance does not improve for $15$ consecutive epochs.

Classification accuracy is used as the primary metric and is reported on the in-distribution, low-SNR, unseen-environment, unseen-communication-parameter, and sea-trial subsets. The simulated OOD average is computed over the low-SNR, unseen-environment, and unseen-communication-parameter results. As an additional external generalization check, we also evaluate selected competitive methods on two public Watermark channel conditions, NOF1 and NCS1. This external test is used only after model selection; no samples from NOF1 or NCS1 are used for training, validation, hyperparameter tuning, or early stopping.

\subsection{Results on UAMR-ShiftBench}\label{subsec:results_overall}

Table~\ref{tab:overall_comparison} reports the benchmark results for SCP-TriCA and representative baselines. SCP-TriCA achieves the best result in all evaluated settings of UAMR-ShiftBench: in-distribution, low-SNR, unseen environment, unseen communication parameters, and the March and November sea-trial subsets. In terms of the main robustness metric, it reaches an OOD average of $74.59\%$, improving over the strongest baseline IQFormer by $5.12$ percentage points. The low R\&CNN accuracy mainly reflects a representation and protocol mismatch: its original design uses long normalized waveform segments reshaped into a $39\times1280$ sequence under a matched sea-trial setting \cite{Zhang2022RCNN}, whereas UAMR-ShiftBench evaluates passband waveform recognition under a broader seven-class, multi-condition protocol.

The performance gap becomes more visible under distribution shift than under matched conditions. While the in-distribution margin over IQFormer is $1.37$ points, the gains increase to $5.22$ points on the low-SNR subset, $4.40$ points under unseen environments, and $5.74$ points under unseen communication parameters. On the March and November sea-trial subsets, SCP-TriCA reaches $91.14\%$ and $94.86\%$, exceeding the best baseline methods by $15.71$ and $23.00$ points, respectively. The low-SNR subset remains the most difficult simulated setting for all methods, but SCP-TriCA still gives the highest accuracy, indicating better robustness when the test SNR falls below the training range.

Fig.~\ref{fig:models_compare_d1_d4} depicts the recognition accuracy of each method across different in-band SNR levels for the four benchmark settings in more detail. In the in-distribution setting, the strongest methods are relatively close. Under the shifted settings, however, the curves separate more clearly, especially in the low-SNR region.  This indicates that the advantage of SCP-TriCA mainly appears under shifted test conditions, particularly when distribution shift and reduced SNR occur together, rather than only under matched conditions.
\begin{figure*}[!t]
\centering
\subfloat[S\&SEFM~\cite{Wang2024FeatureFusion}.]{%
\begin{minipage}[t]{0.27\linewidth}
\centering
\includegraphics[width=\linewidth]{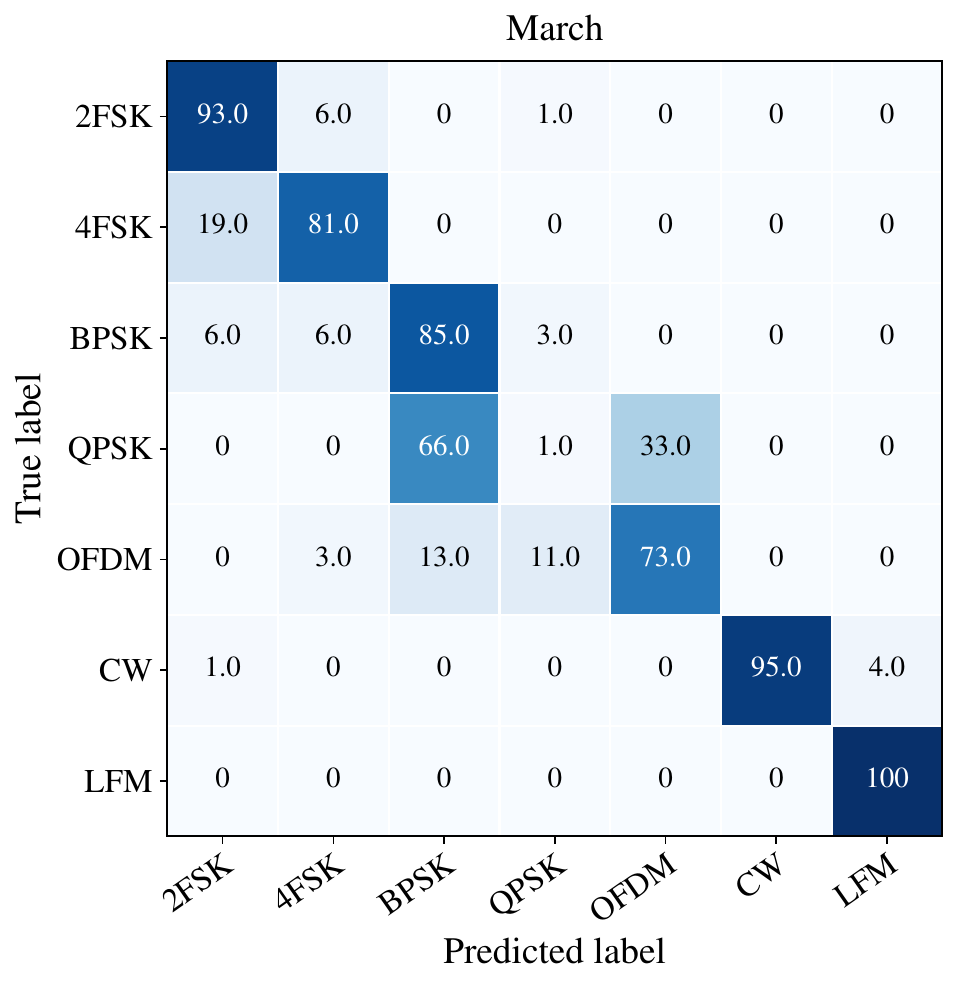}
\vspace{0.04em}
\includegraphics[width=\linewidth]{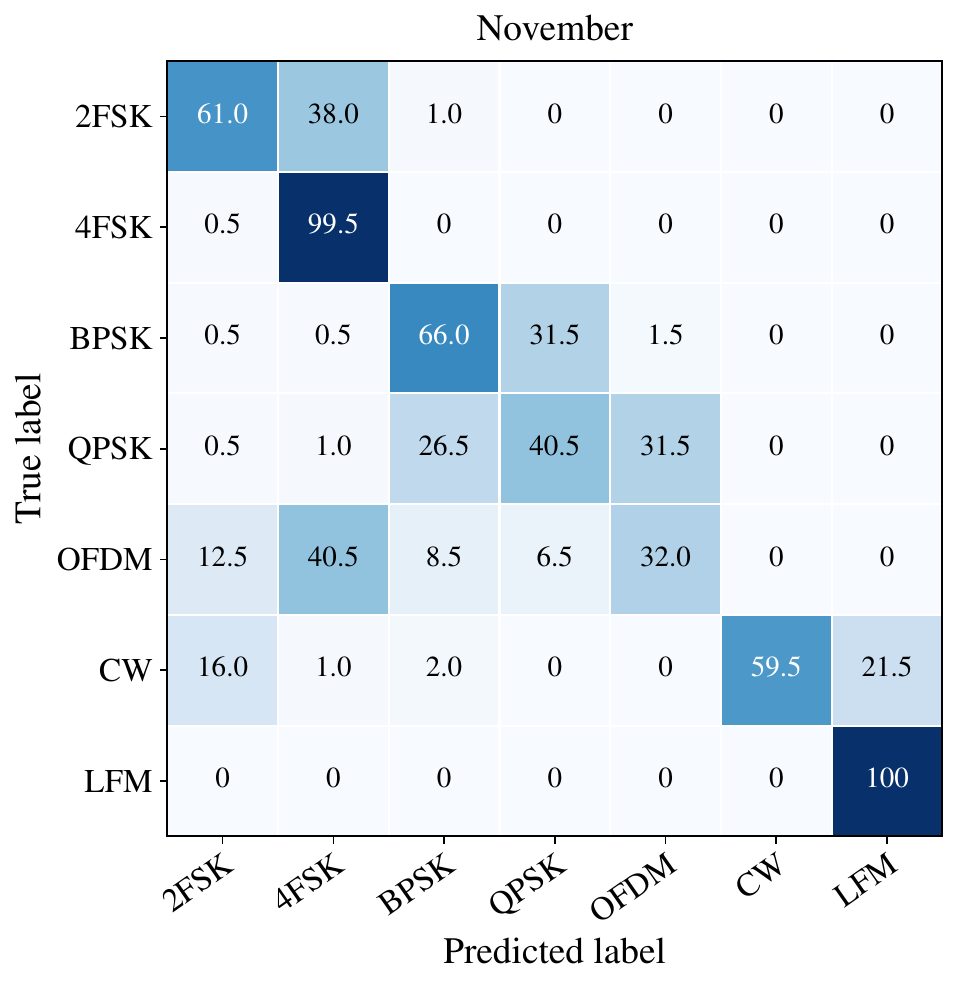}
\end{minipage}}
\hspace{0.2em}
\subfloat[IQFormer~\cite{Shao2025IQFormer}.]{%
\begin{minipage}[t]{0.27\linewidth}
\centering
\includegraphics[width=\linewidth]{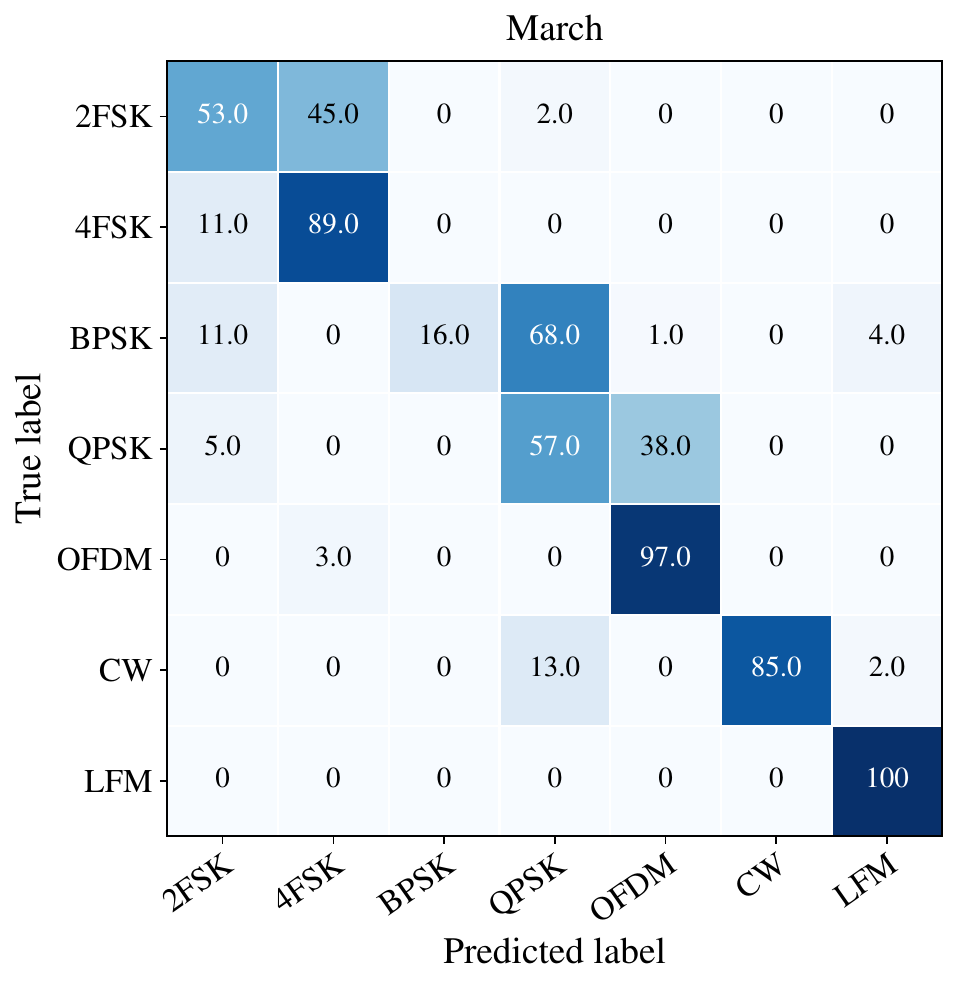}
\vspace{0.04em}
\includegraphics[width=\linewidth]{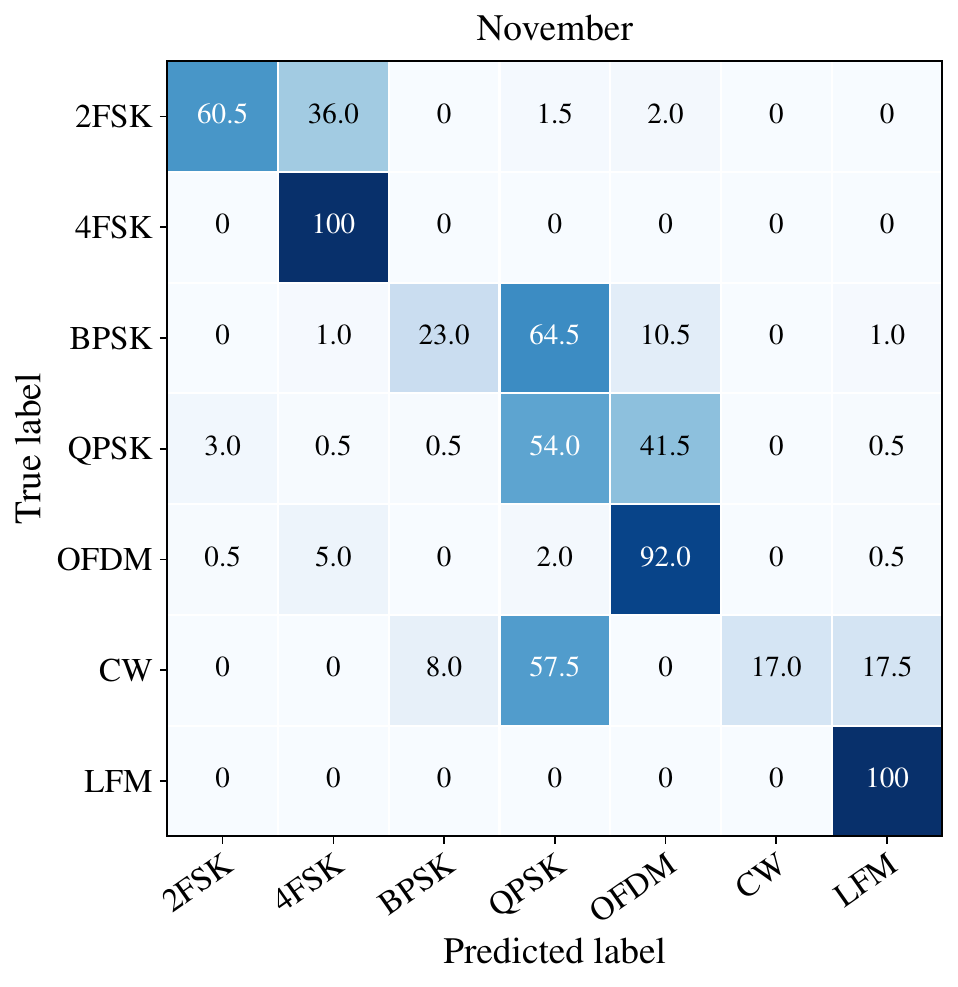}
\end{minipage}}
\hspace{0.2em}
\subfloat[SCP-TriCA (ours).]{%
\begin{minipage}[t]{0.27\linewidth}
\centering
\includegraphics[width=\linewidth]{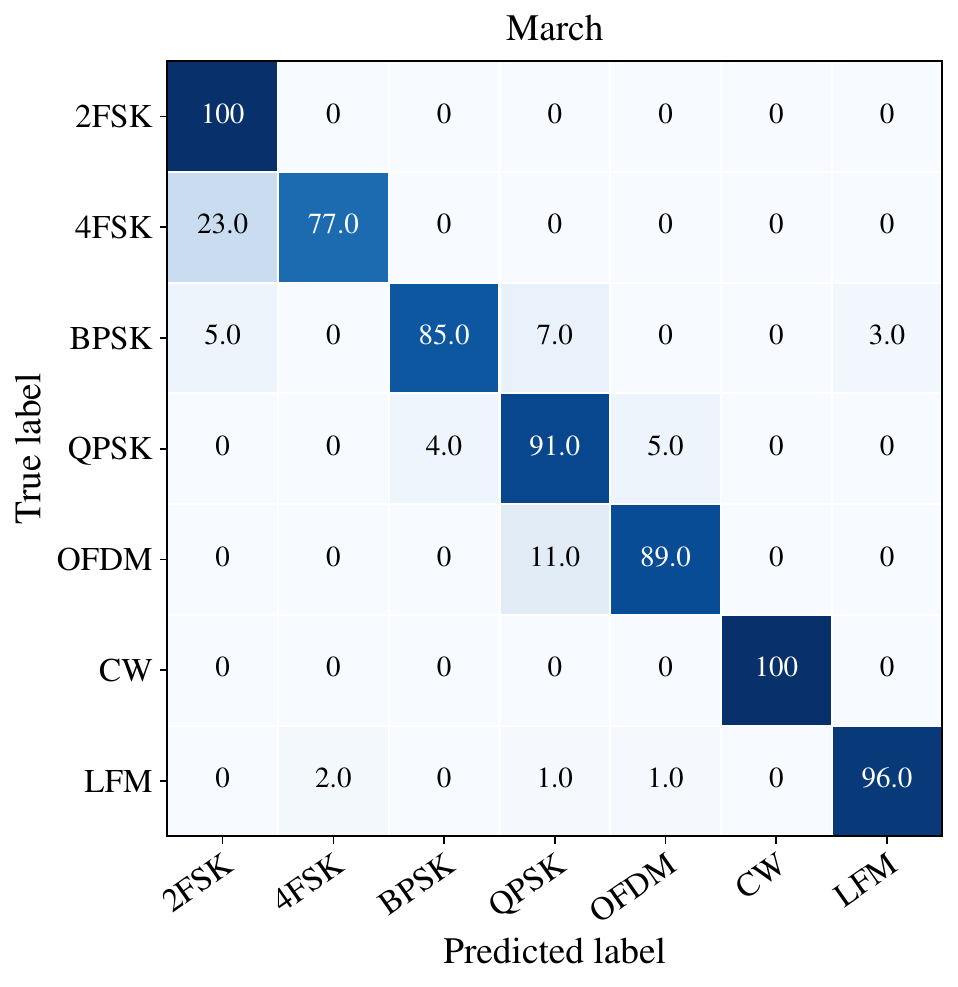}
\vspace{0.04em}
\includegraphics[width=\linewidth]{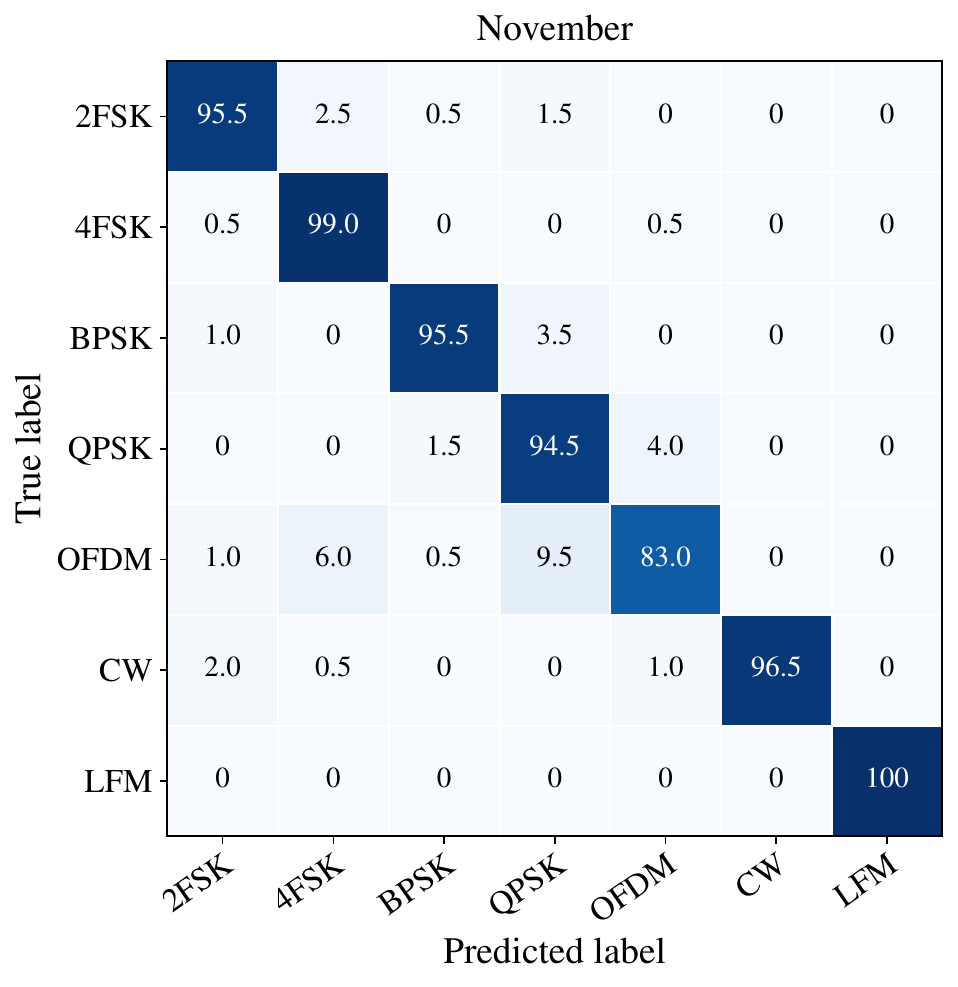}
\end{minipage}}
\caption{Sea-trial confusion-matrix comparison of S\&SEFM, IQFormer, and SCP-TriCA on UAMR-ShiftBench. }
\label{fig:seatrial_confusion_matrices}
\end{figure*}

Fig.~\ref{fig:seatrial_confusion_matrices} compares the sea-trial confusion matrices of SCP-TriCA with two representative strong baselines, S\&SEFM and IQFormer. Consistent with Table~\ref{tab:overall_comparison}, SCP-TriCA yields a much clearer diagonal structure on both the March and November subsets, reaching $91.14\%$ and $94.86\%$ accuracy, respectively. The remaining errors are mainly associated with phase-modulated and OFDM-related categories, with some residual confusion within the FSK family. These patterns suggest that the measured sea-trial channel still weakens class-discriminative cues for closely related waveforms, but SCP-TriCA suppresses such off-diagonal errors more effectively than the baselines.

\begin{table}[h]
\caption{Evaluation on public Watermark channel conditions.}
\label{tab:watermark_external}
\centering
\footnotesize
\renewcommand{\arraystretch}{1.15}
\setlength{\tabcolsep}{5pt}
\begin{tabular}{lccc}
\toprule
Method & \tablehead{NOF1 (\%)} & \tablehead{NCS1 (\%)} & \tablehead{Average (\%)} \\
\midrule
TSTR~\cite{Li2024TSTR} & 77.86 & 69.56 & 73.71 \\
S\&SEFM~\cite{Wang2024FeatureFusion} & 78.18 & 65.35 & 71.77 \\
IQFormer~\cite{Shao2025IQFormer} & \underline{91.94} & \underline{70.34} & \underline{81.14} \\
\midrule
\rowcolor{TableOurs}
SCP-TriCA (ours) & \textbf{94.14} & \textbf{90.34} & \textbf{92.24} \\
\bottomrule
\end{tabular}
\end{table}
\vspace{-1 em}
\subsection{Results on Watermark Channels}\label{subsec:results_watermark}

Beyond the shift-disentangled evaluation provided by UAMR-ShiftBench, an additional concern is that a model might implicitly adapt to the specific channel-generation and evaluation choices used by the proposed benchmark. To examine this issue, we construct an extra test-only set by passing the same seven waveform classes through two public Watermark channel conditions, NOF1 and NCS1~\cite{VanWalree2017Watermark}. NOF1 represents a shallow Oslofjord channel with relatively stable arrivals, whereas NCS1 is measured on the Norwegian continental shelf and exhibits stronger time variation and less stable multipath structure, making it the more challenging of the two channels. Following the frequency-band requirement of these Watermark channels, the carrier or center frequency is restricted to 10--18 kHz, the in-band SNR is set to 0--20 dB, and each channel contains 14,000 test samples. 

Table~\ref{tab:watermark_external} reports the zero-shot results obtained by directly applying the same trained checkpoints from Table~\ref{tab:overall_comparison} to the NOF1 and NCS1 test sets. SCP-TriCA achieves the highest accuracy on both Watermark channels, with an average accuracy of $92.24\%$. Compared with the strongest baseline IQFormer, SCP-TriCA improves the average result by $11.10$ percentage points. The gain is especially pronounced on NCS1, where the stronger time variation and less stable multipath structure cause larger degradation for the baselines, whereas SCP-TriCA maintains $90.34\%$ accuracy. These results indicate that SCP-TriCA exhibits better cross-channel generalization than the compared methods under channel conditions outside UAMR-ShiftBench.

\begin{table*}[t]
\caption{Main ablation results of \textbf{SCP-TriCA}. The best results are highlighted in \textbf{bold}, and the second-best results are \underline{underscored}.}
\label{tab:ablation_main}
\centering
\small
\setlength{\tabcolsep}{4pt}
\resizebox{\textwidth}{!}{%
\begin{tabular}{lcccccccccc}
\toprule
Variant & \tablehead{STFT} & \tablehead{Cyc} & \tablehead{P2/P4} & \tablehead{Interaction\\design} & \tablehead{In-Distribution\\(\%)} & \tablehead{Low-SNR\\(\%)} & \tablehead{Unseen\\Environment (\%)} & \tablehead{Unseen Comm.\\Parameters (\%)} & \tablehead{March\\Sea-Trial (\%)} & \tablehead{November\\Sea-Trial (\%)} \\
\midrule
\rowcolor{TableMetric}
\multicolumn{11}{c}{\emph{Input-modality ablation}} \\
\midrule
STFT only & \cmark & \xmark & \xmark & Single branch & 86.99 & 43.05 & 63.32 & 63.26 & 76.29 & 77.21 \\
Cyc only & \xmark & \cmark & \xmark & Single branch & 82.97 & 62.13 & 72.95 & 68.05 & 65.86 & 64.57\\
P2/P4 only & \xmark & \xmark & \cmark & Single branch & 84.21 & 39.79 & 61.44 & 60.73 & 80.86 & 67.07 \\
STFT + Cyc & \cmark & \cmark & \xmark & \makecell{Stage 1 only:\\bidirectional 2D CA} & 89.08 & \underline{64.29} & 76.12 & 75.30 & 83.71 & 79.14 \\
\midrule
\rowcolor{TableMetric}
\multicolumn{11}{c}{\emph{Fusion-strategy ablation}} \\
\midrule
Direct CA & \cmark & \cmark & \cmark & \makecell{Direct simultaneous\\tri-modal cross-attention} & \underline{95.16} & 64.24 & \underline{78.66} & 78.36 & 86.86 & 84.79 \\
Early concat & \cmark & \cmark & \cmark & Front-end concat & 86.37 & 62.08 & 73.84 & 73.79 & 79.00 & 85.43 \\
Late concat & \cmark & \cmark & \cmark & Post-encoder concat & 94.41 & 63.55 & 78.15 & 77.31 & 85.29 & \underline{94.07} \\
w/o selective gate & \cmark & \cmark & \cmark & \makecell{Ungated P2/P4\\injection} & 94.84 & 63.62 & 78.30 & \underline{78.45} & \underline{90.86} & 78.14 \\
\midrule
\rowcolor{TableOurs}
SCP-TriCA (ours) & \cmark & \cmark & \cmark & \makecell{2D$\leftrightarrow$2D first,\\then gated P2/P4 injection} & \textbf{95.33} & \textbf{65.48} & \textbf{79.06} & \textbf{79.23} & \textbf{91.14} & \textbf{94.86} \\
\bottomrule
\end{tabular}}
\end{table*}

\begin{figure*}[!t]
\centering
\begin{minipage}[t]{0.48\textwidth}
\centering
\subfloat[]{%
\includegraphics[width=0.48\linewidth]{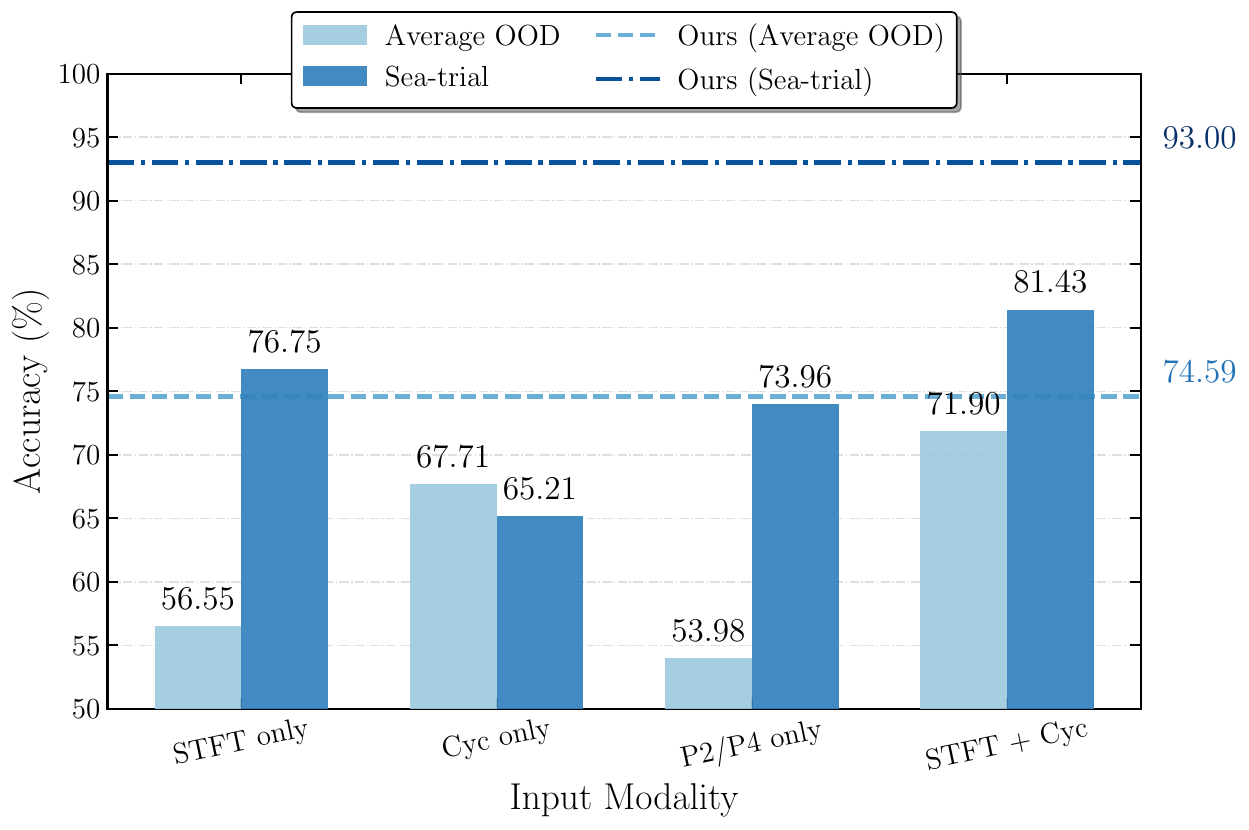}}
\hfil
\subfloat[]{%
\includegraphics[width=0.48\linewidth]{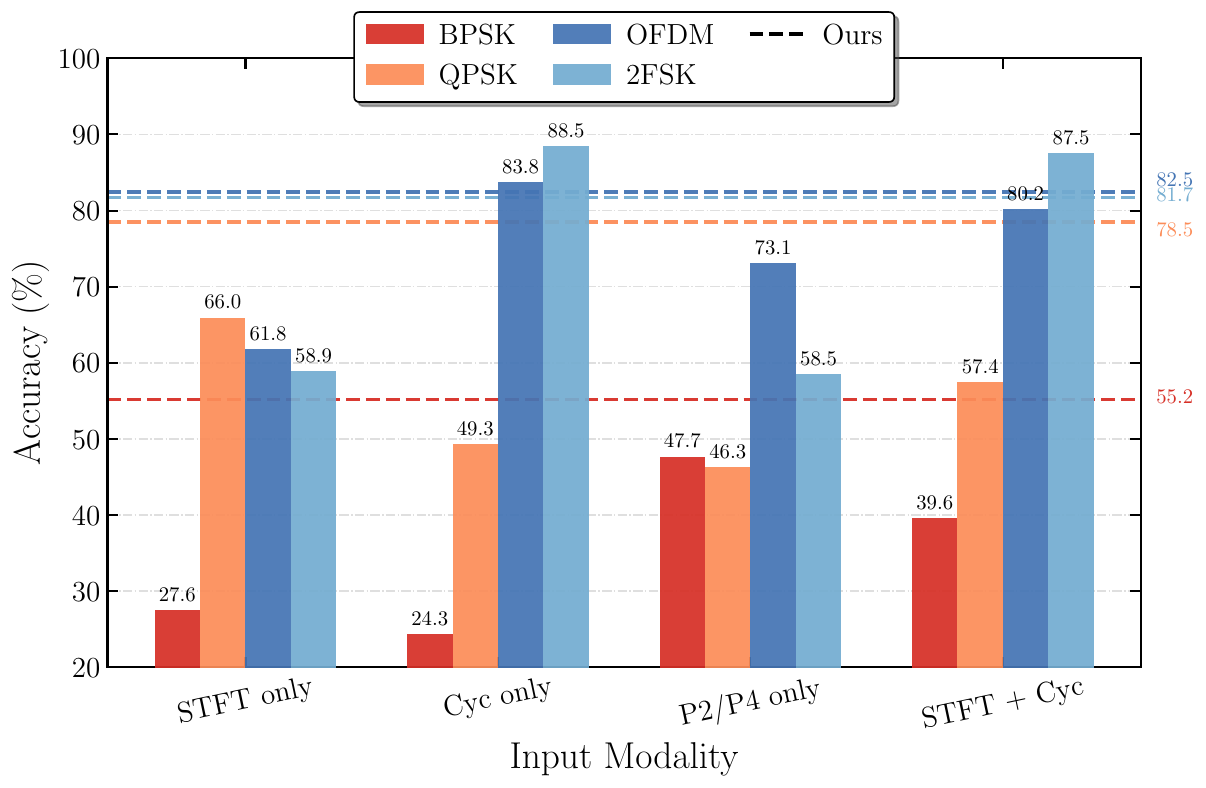}}
\caption{Input-modality ablation of SCP-TriCA. (a) Average OOD and sea-trial average accuracy. (b) Challenging-category OOD accuracy.}
\label{fig:ablation_input_visualization}
\end{minipage}
\hfill
\begin{minipage}[t]{0.48\textwidth}
\centering
\subfloat[]{%
\includegraphics[width=0.48\linewidth]{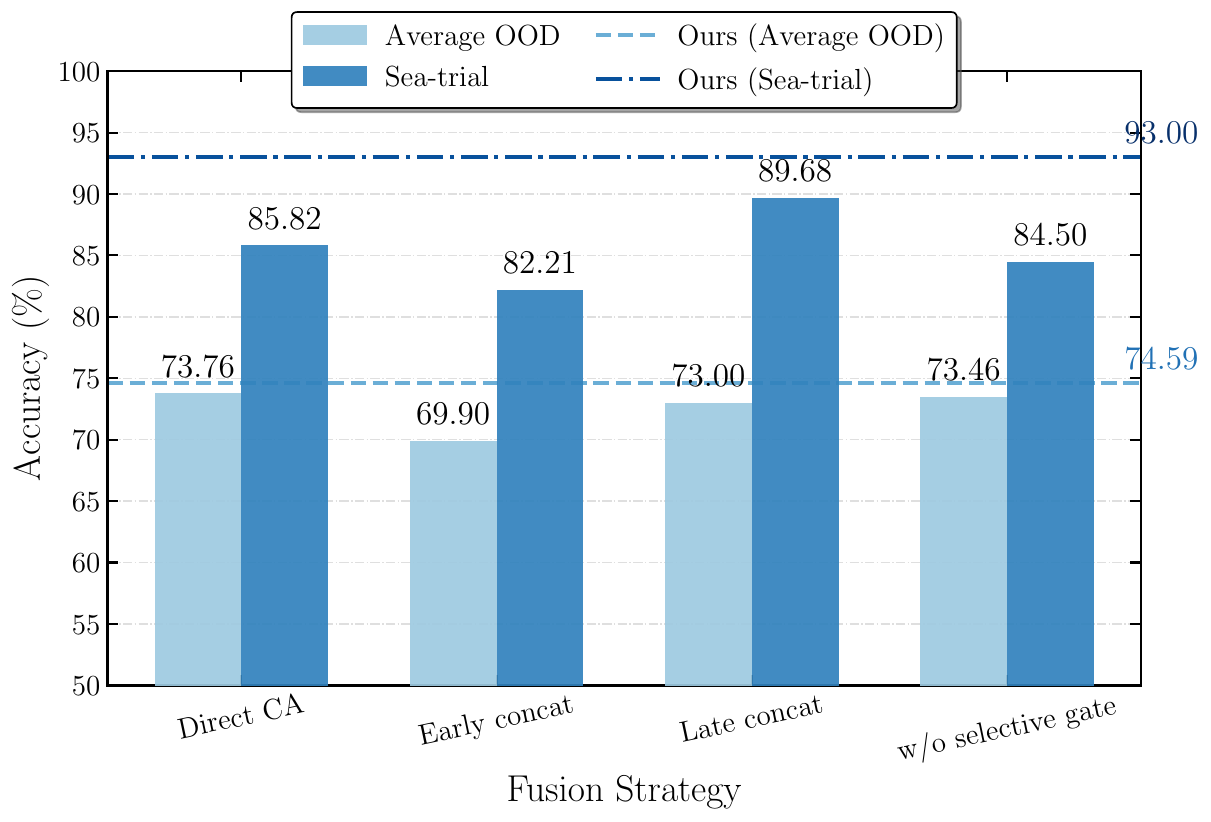}}
\hfil
\subfloat[]{%
\includegraphics[width=0.48\linewidth]{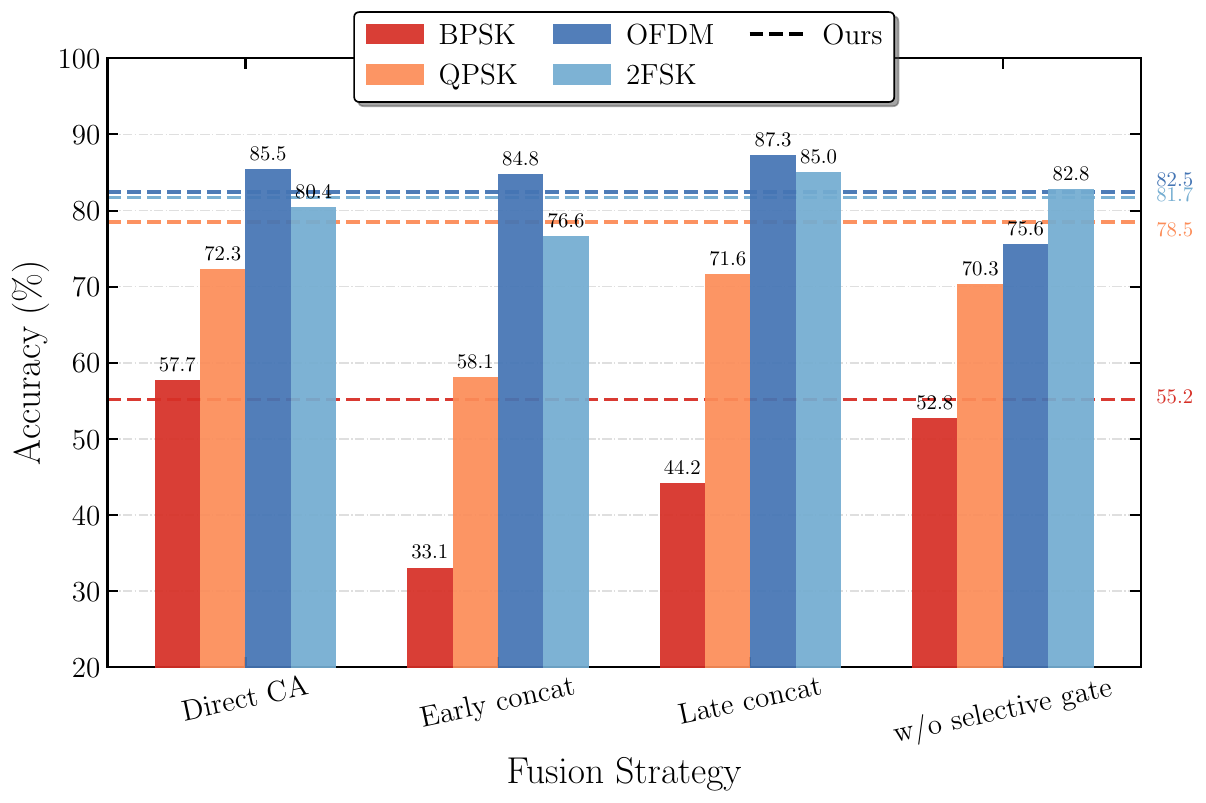}}
\caption{Fusion-strategy ablation of SCP-TriCA. (a) Average OOD and sea-trial average accuracy. (b) Challenging-category OOD accuracy.}
\label{fig:ablation_fusion_visualization}
\end{minipage}
\end{figure*}

\subsection{Ablation Study}\label{subsec:results_ablation}

An ablation study was conducted under the same UAMR-ShiftBench protocol to determine the benefits of the main components of SCP-TriCA. Table~\ref{tab:ablation_main} summarizes the input-modality ablation and fusion-strategy ablation. The results show that each modality contributes to the final performance, while the hierarchical fusion order and selective gate provide additional gains.

The input-modality ablation reveals a clear division of labor among the three modalities, as summarized in Table~\ref{tab:ablation_main} and Fig.~\ref{fig:ablation_input_visualization}. Among the single-modality variants, the Cyc-only variant is the most robust under simulated shifts, reaching $67.71\%$ average OOD, whereas the STFT-only variant gives stronger in-distribution recognition but degrades more under low SNR. Combining STFT and Cyc already captures much of the transferable 2D evidence, improving the average OOD to $71.90\%$ and the sea-trial average to $81.43\%$. Adding P2/P4 through the gated tri-modal design further raises these results to $74.59\%$ and $93.00\%$, respectively. This supports the view that P2/P4 is most useful as complementary, sample-adaptive statistical evidence rather than as a standalone branch, especially for measured-data transfer.

The fusion-strategy ablation shows that, once all three modalities are present, overall robustness depends strongly on how the modalities interact, as shown in Fig.~\ref{fig:ablation_fusion_visualization}. Early concatenation is the weakest tri-modal design on simulated OOD, reaching only $69.90\%$ average OOD and $82.21\%$ sea-trial average accuracy, indicating that forcing heterogeneous cues into one shared representation too early still limits robustness under simulated distribution shifts. Late concatenation improves the sea-trial average to $89.68\%$, but its simulated OOD average remains $73.00\%$, suggesting that simply merging independently encoded branches can transfer well to the measured November subset but is still less balanced than the proposed design across simulated shifts. Direct tri-modal cross-attention gives the strongest simulated OOD result among the alternative fusion strategies ($73.76\%$), but its sea-trial average remains $85.82\%$, indicating that unrestricted simultaneous interaction is not sufficient for reliable simulation-to-real transfer.

The ungated hierarchical variant keeps the same fusion order as SCP-TriCA but removes the adaptive P2/P4 gate. Its simulated OOD average remains close to SCP-TriCA ($73.46\%$ versus $74.59\%$), whereas the sea-trial average drops from $93.00\%$ to $84.50\%$, mainly due to the November subset. This pattern is consistent with the single-modality results: P2/P4 alone is more effective on the March subset than on the November subset ($80.86\%$ versus $67.07\%$), so ungated injection is tolerated in March but becomes harmful when the statistical cues are less reliable.  Thus, the hierarchy contributes most of the simulated-shift robustness, while the gate is critical for measured-data transfer. SCP-TriCA achieves the best joint result, with $74.59\%$ average OOD and $93.00\%$ sea-trial average accuracy, and Fig.~\ref{fig:ablation_fusion_visualization}(b) shows a more balanced profile across difficult waveform categories.

\begin{figure*}[t]
\centering
\setlength{\gateDiagHeight}{0.16\textwidth}
\begin{minipage}[t]{0.30\textwidth}
\centering
\includegraphics[height=\gateDiagHeight,trim=9.4pt 9.4pt 5.5pt 28.7pt,clip]{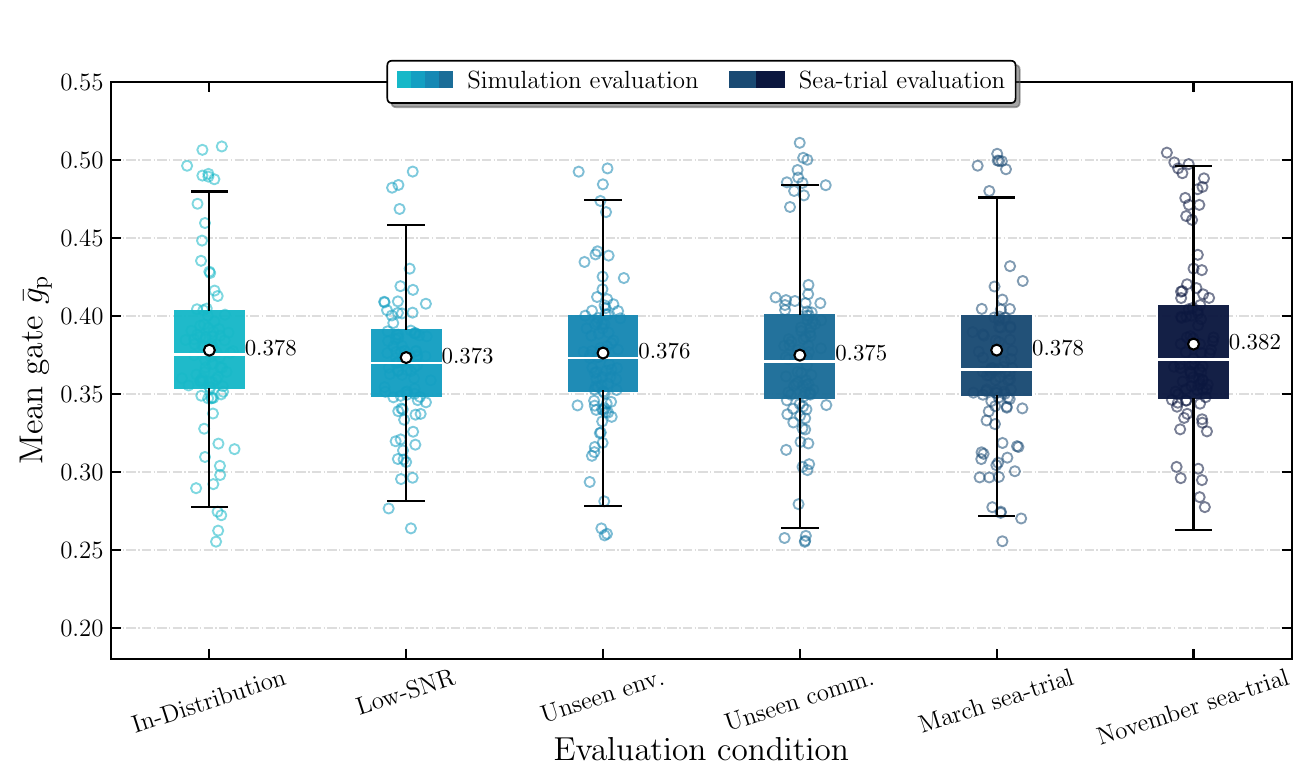}
\par\smallskip
{\normalfont\normalsize (a)\par}
\end{minipage}
\hfil
\begin{minipage}[t]{0.30\textwidth}
\centering
\includegraphics[height=\gateDiagHeight,trim=9.4pt 9.2pt 5.5pt 28.7pt,clip]{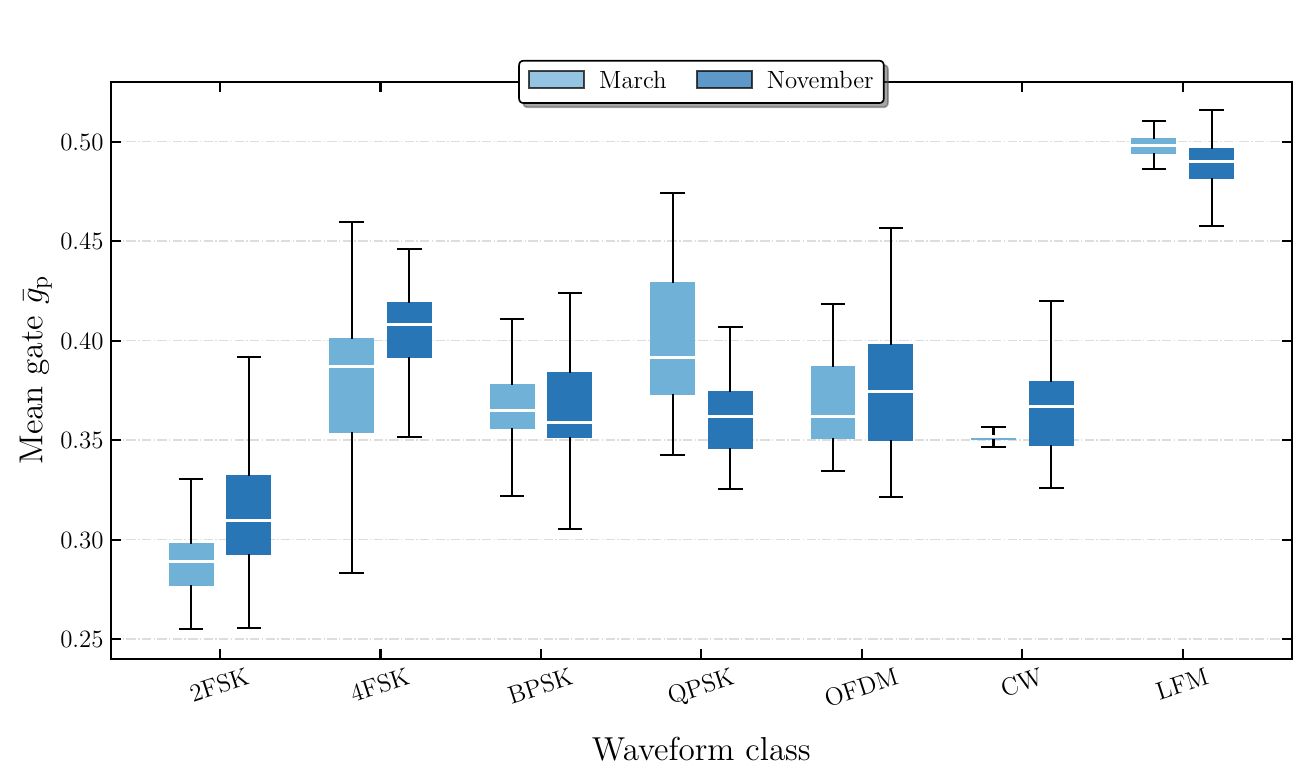}
\par\smallskip
{\normalfont\normalsize (b)\par}
\end{minipage}
\hfil
\begin{minipage}[t]{0.34\textwidth}
\centering
\includegraphics[height=\gateDiagHeight,trim=4.2pt 7.1pt 4.3pt 3.9pt,clip]{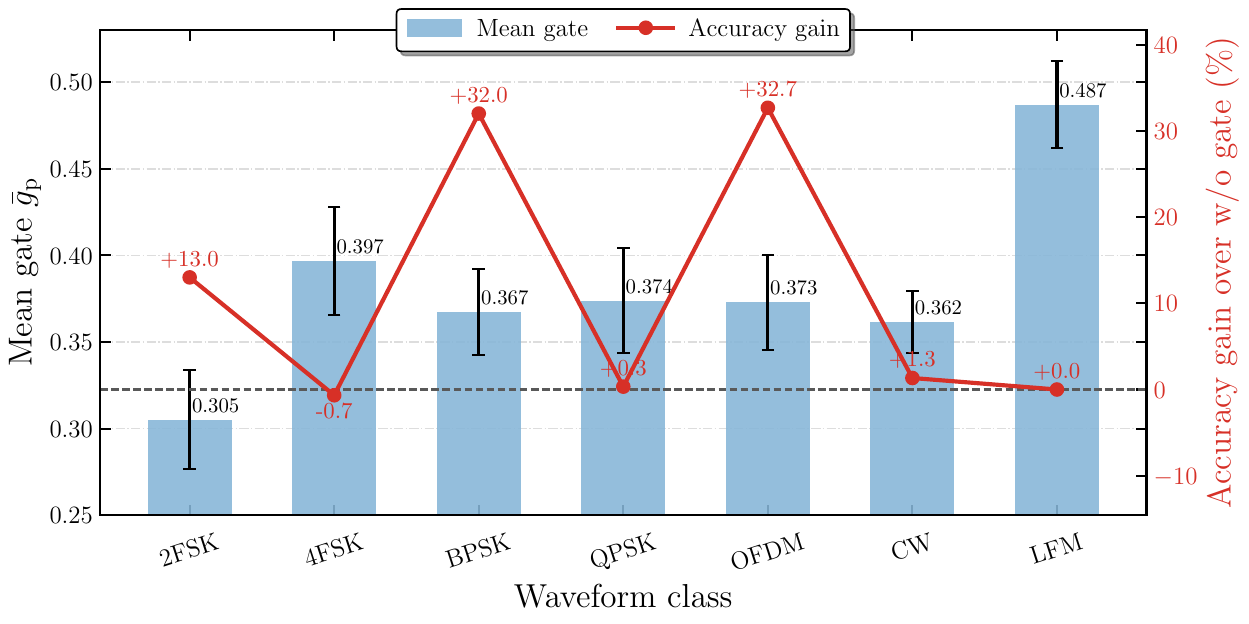}
\par\smallskip
{\normalfont\normalsize (c)\par}
\end{minipage}
\caption{Analysis of the sample-adaptive P2/P4 injection gate. (a) Gate distributions across evaluation settings. (b) Sea-trial class distributions. (c) Class-wise mean gate values over all March and November sea-trial samples and the paired accuracy gain relative to the ungated hierarchical variant.}
\label{fig:gate_value_diagnostics}
\end{figure*}

To further inspect the selective gate, Fig.~\ref{fig:gate_value_diagnostics} visualizes the sample-level mean gate $\bar{g}_{\mathrm{p}}=\frac{1}{d_p}\sum_{c=1}^{d_p}g_{\mathrm{p},c}$, where lower values preserve the fused 2D representation and higher values inject more P2/P4-updated evidence. The gate values remain in a moderate range across evaluation settings, indicating that the model does not simply switch the statistical branch on or off by test condition. Instead, the distributions vary across sea-trial waveform classes, and the paired accuracy gain over the ungated variant is not monotonic with the absolute gate magnitude. The largest gains occur for BPSK and OFDM ($+32.0$ and $+32.7$ points), whereas QPSK is nearly neutral ($+0.3$ points). These diagnostics support the ablation results: the gate improves measured-data transfer by selectively regulating P2/P4 refinement according to the sample and class.

\section{Conclusion}\label{sec:conclusion}

This paper used underwater acoustic modulation recognition to examine a broader robustness problem: how to fuse signal-image and statistical representations when their reliability changes across operating conditions. To support this study, we constructed UAMR-ShiftBench and proposed SCP-TriCA, a tri-modal cross-attention framework that integrates STFT, cyclostationary, and P2/P4 modalities through a hierarchical fusion strategy. Experimental results show that SCP-TriCA achieves the strongest overall performance under the evaluated shifted conditions, especially in tests with low SNR, unseen environments, unseen communication parameters, and measured sea-trial signals. These results indicate better robustness and generalization than the compared baselines. The ablation study further confirms that these gains come from modality complementarity and from organizing heterogeneous 2D signal-image and 1D statistical evidence according to their different reliability under distribution shift.

Beyond these results, UAMR-ShiftBench provides a standardized basis for robustness evaluation, enabling future models to be compared under consistent criteria and helping researchers identify whether performance degradation comes from noise, environmental mismatch, parameter variation, or simulation-to-real transfer. Meanwhile, the design of SCP-TriCA provides a practical reference for organizing heterogeneous signal representations in related underwater monitoring and non-cooperative communication-analysis tasks. Future work will extend the benchmark with more diverse sea-trial measurements, broader waveform classes, and more realistic deployment scenarios, and will further improve SCP-TriCA by incorporating temporal and multihydrophone evidence, enhancing sample-adaptive modality reliability estimation, and developing lightweight variants for practical underwater acoustic recognition systems.

\appendices
\begin{table}[!t]
\caption{Signal-specific parameter ranges used in UAMR-ShiftBench}
\label{tab:params}
\centering
\footnotesize
\renewcommand{\arraystretch}{1.0}
\setlength{\tabcolsep}{2pt}
\begin{tabular}{@{}>{\raggedright\arraybackslash}p{0.15\columnwidth}>{\raggedright\arraybackslash}p{0.84\columnwidth}@{}}
\toprule
Signal type & Parameter settings \\
\midrule
2FSK & Symbol rate $\mathcal{R}_s$: 25--800 Bd; modulation index $h \in \{2, 3, 4, 5\}$ \\
4FSK & Symbol rate $\mathcal{R}_s$: 20--400 Bd; modulation index $h \in \{2, 3, 4, 5\}$ \\
BPSK & Symbol rate $\mathcal{R}_s$: 40--1000 Bd; raised-cosine shaping is used with probability 0.8 and a non-raised-cosine setting with probability 0.2; roll-off factor $\mathcal{\beta}$ $\in \{0.50, 0.60, 0.70, 0.80\}$ for raised-cosine cases \\
QPSK & Same parameterization as BPSK \\
OFDM & Occupied bandwidth $\mathcal{B}$: 500--3200 Hz; number of subcarriers $\in \{512, 1024, 2048\}$; cyclic-prefix ratio $CP \in \{0, 0.125\}$, with $CP=0.125$ used with probability 0.25 \\
CW & Narrowband single-tone signals \\
LFM & Sweep bandwidth $\mathcal{B}$: 80--2000 Hz \\
\bottomrule
\end{tabular}
\end{table}

\begin{table}[!t]
\caption{Environment and data-split settings of UAMR-ShiftBench.}
\label{tab:app_env_split}
\centering
\footnotesize
\renewcommand{\arraystretch}{1}
\setlength{\tabcolsep}{2pt}
\begin{tabular}{@{}>{\raggedright\arraybackslash}p{0.36\columnwidth}@{\hspace{0.02\columnwidth}}>{\raggedright\arraybackslash}p{0.62\columnwidth}@{}}
\toprule
Item & Configuration \\
\midrule
Simulation locations &
\begin{tabular}[t]{@{}ll@{}}
Location 1: & $(18.75^\circ\mathrm{N}, 107.9^\circ\mathrm{E})$; \\
Location 2: & $(17.4^\circ\mathrm{N}, 108.7^\circ\mathrm{E})$; \\
Location 3: & $(19.8^\circ\mathrm{N}, 112.3^\circ\mathrm{E})$
\end{tabular} \\
Bottom configuration & Flat bottom; homogeneous fluid half-space with $c_b=1549.5~\mathrm{m/s}$, $\rho_b=1.583~\mathrm{g/cm^3}$, and $\alpha_b=0.11~\mathrm{dB}/\lambda$ \\
Propagation range & $[8,35]$ km \\
Source and receiver depths & Two source depths are used: a near-surface source at 10 m and a submerged source at 30 m; receiver depth is sampled from 40--50 m \\
Frequency-band partition & Bands 1--8: $[0.8,1.5]$, $[1.5,2.5]$, $[2.5,4]$, $[4,6]$, $[6,9]$, $[9,13]$, $[13,19]$, and $[19,25]$ kHz \\
Matched-condition bands & Frequency-band indices [1, 3, 5, 7, 8] \\
Held-out bands & Frequency-band indices [2, 4, 6] for signal-parameter shift \\
Held-out symbol rates & Held-out discrete baud-rate values that do not appear in the training split for signal-parameter shift \\

\bottomrule
\end{tabular}
\end{table}

\vspace{-1em}

\section{Dataset Configuration Details}\label{app:dataset_config}\label{appsubsec:dataset_env}
This appendix provides the detailed dataset configuration supporting the benchmark overview in Section~\ref{subsec:exp_dataset}. Table~\ref{tab:params} lists the class-dependent signal-parameter ranges for the seven waveform classes. Table~\ref{tab:app_env_split} reports the simulation-environment settings, the BELLHOP-based propagation configuration used to generate simulated channels~\cite{An2022LowResolutionFourier,PorterBucker1987Bellhop}, and the frequency-band and symbol-rate partitions used for the signal-parameter shift. The simulated propagation range is set to 8--35 km, corresponding to the medium-range shallow-water setting considered in this benchmark. The two source depths represent typical shallow-water deployment geometries: a near-surface source at 10 m and a submerged source at 30 m. The receiver depth is randomized between 40 and 50 m to introduce moderate source--receiver geometry variation.

\bibliographystyle{IEEEtran}
\bibliography{mypaper-ref}

\end{document}